\pdfoutput=1
\documentclass[twocolumn,numberedappendix]{aastex701}
\usepackage{amsmath}
\usepackage{amssymb}
\usepackage{natbib}
\usepackage{graphicx} %
\usepackage{color}
\usepackage{bm}
\usepackage{float}
\usepackage{multirow}
\usepackage{booktabs}
\usepackage{placeins}
\usepackage{rotating}
\usepackage{chngcntr}
\usepackage{savesym}
\savesymbol{tablenum}
\usepackage[utf8]{inputenc}
\usepackage{siunitx}
\restoresymbol{SIX}{tablenum}

\DeclareSIUnit{\erg}{erg}

\newcommand{\ignore}[1]{}
\makeatletter\def\ApjSectionpenalty{\@M}\makeatother
\usepackage{hyperref} %

\shorttitle{Extragalactic Transients from SPT-3G}
\shortauthors{Guns et al.}

\begin{document}

\bibliographystyle{yahapj}

\title{Constraints on Millimeter-wave Extragalactic Transient Populations from 4 Years of SPT-3G data}

\author[0000-0001-7143-2853]{S.~Guns}
\affiliation{Department of Physics, University of California, Berkeley, CA, 94720, USA}
\email{sguns@berkeley.edu}

\author[0000-0002-2077-6004]{C.~Tandoi}
\affiliation{Department of Astronomy, University of Illinois Urbana-Champaign, 1002 West Green Street, Urbana, IL, 61801, USA}
\email{ctandoi2@illinois.edu}

\author{D.~Anbajagane}
\affiliation{Department of Astronomy and Astrophysics, University of Chicago, 5640 South Ellis Avenue, Chicago, IL, 60637, USA}
\affiliation{Kavli Institute for Cosmological Physics, University of Chicago, 5640 South Ellis Avenue, Chicago, IL, 60637, USA}
\email{dhayaa@uchicago.edu}

\author[0009-0003-3245-3979]{E.~Anderes}
\affiliation{Department of Statistics, University of California, One Shields Avenue, Davis, CA 95616, USA}
\email{ebanderes@ucdavis.edu}

\author[0000-0002-4435-4623]{A.~J.~Anderson}
\affiliation{Fermi National Accelerator Laboratory, MS209, P.O. Box 500, Batavia, IL, 60510, USA}
\affiliation{Kavli Institute for Cosmological Physics, University of Chicago, 5640 South Ellis Avenue, Chicago, IL, 60637, USA}
\affiliation{Department of Astronomy and Astrophysics, University of Chicago, 5640 South Ellis Avenue, Chicago, IL, 60637, USA}
\email{adama@fnal.gov}

\author[0000-0002-6443-3396]{B.~Ansarinejad}
\affiliation{School of Physics, University of Melbourne, Parkville, VIC 3010, Australia}
\email{behzad.ansarinejad@gmail.com}

\author[0000-0002-0517-9842]{M.~Archipley}
\affiliation{Department of Astronomy and Astrophysics, University of Chicago, 5640 South Ellis Avenue, Chicago, IL, 60637, USA}
\affiliation{Kavli Institute for Cosmological Physics, University of Chicago, 5640 South Ellis Avenue, Chicago, IL, 60637, USA}
\email{archipleym@uchicago.edu}

\author[0000-0001-6899-1873]{L.~Balkenhol}
\affiliation{Sorbonne Universit\'e, CNRS, UMR 7095, Institut d'Astrophysique de Paris, 98 bis bd Arago, 75014 Paris, France}
\email{lennart.balkenhol@iap.fr}

\author[0000-0002-1623-5651]{D.~R.~Barron}
\affiliation{Department of Physics and Astronomy, University of New Mexico, Albuquerque, NM, 87131, USA}
\email{barrondarcy@gmail.com}

\author{K.~Benabed}
\affiliation{Sorbonne Universit\'e, CNRS, UMR 7095, Institut d'Astrophysique de Paris, 98 bis bd Arago, 75014 Paris, France}
\email{benabed@iap.fr}

\author[0000-0001-5868-0748]{A.~N.~Bender}
\affiliation{High-Energy Physics Division, Argonne National Laboratory, 9700 South Cass Avenue, Lemont, IL, 60439, USA}
\affiliation{Kavli Institute for Cosmological Physics, University of Chicago, 5640 South Ellis Avenue, Chicago, IL, 60637, USA}
\affiliation{Department of Astronomy and Astrophysics, University of Chicago, 5640 South Ellis Avenue, Chicago, IL, 60637, USA}
\email{abender@anl.gov}

\author[0000-0002-5108-6823]{B.~A.~Benson}
\affiliation{Fermi National Accelerator Laboratory, MS209, P.O. Box 500, Batavia, IL, 60510, USA}
\affiliation{Kavli Institute for Cosmological Physics, University of Chicago, 5640 South Ellis Avenue, Chicago, IL, 60637, USA}
\affiliation{Department of Astronomy and Astrophysics, University of Chicago, 5640 South Ellis Avenue, Chicago, IL, 60637, USA}
\email{bbenson@astro.uchicago.edu}

\author[0000-0003-4847-3483]{F.~Bianchini}
\affiliation{Kavli Institute for Particle Astrophysics and Cosmology, Stanford University, 452 Lomita Mall, Stanford, CA, 94305, USA}
\affiliation{Department of Physics, Stanford University, 382 Via Pueblo Mall, Stanford, CA, 94305, USA}
\affiliation{SLAC National Accelerator Laboratory, 2575 Sand Hill Road, Menlo Park, CA, 94025, USA}
\email{fbianc@stanford.edu}

\author[0000-0001-7665-5079]{L.~E.~Bleem}
\affiliation{High-Energy Physics Division, Argonne National Laboratory, 9700 South Cass Avenue, Lemont, IL, 60439, USA}
\affiliation{Kavli Institute for Cosmological Physics, University of Chicago, 5640 South Ellis Avenue, Chicago, IL, 60637, USA}
\affiliation{Department of Astronomy and Astrophysics, University of Chicago, 5640 South Ellis Avenue, Chicago, IL, 60637, USA}
\email{lbleem@anl.gov}

\author[0000-0002-4900-805X]{S.~Bocquet}
\affiliation{University Observatory, Faculty of Physics, LMU Munich, Scheinerstr.~1, 81679 Munich, Germany}
\email{sebastian.bocquet@physik.lmu.de}

\author[0000-0002-8051-2924]{F.~R.~Bouchet}
\affiliation{Sorbonne Universit\'e, CNRS, UMR 7095, Institut d'Astrophysique de Paris, 98 bis bd Arago, 75014 Paris, France}
\email{bouchet@iap.fr}

\author[0000-0003-3483-8461]{E.~Camphuis}
\affiliation{Sorbonne Universit\'e, CNRS, UMR 7095, Institut d'Astrophysique de Paris, 98 bis bd Arago, 75014 Paris, France}
\email{etienne.camphuis@iap.fr}

\author{M.~G.~Campitiello}
\affiliation{High-Energy Physics Division, Argonne National Laboratory, 9700 South Cass Avenue, Lemont, IL, 60439, USA}
\email{mcampitiello@anl.gov}

\author[0000-0002-2044-7665]{J.~E.~Carlstrom}
\affiliation{Kavli Institute for Cosmological Physics, University of Chicago, 5640 South Ellis Avenue, Chicago, IL, 60637, USA}
\affiliation{Enrico Fermi Institute, University of Chicago, 5640 South Ellis Avenue, Chicago, IL, 60637, USA}
\affiliation{Department of Physics, University of Chicago, 5640 South Ellis Avenue, Chicago, IL, 60637, USA}
\affiliation{High-Energy Physics Division, Argonne National Laboratory, 9700 South Cass Avenue, Lemont, IL, 60439, USA}
\affiliation{Department of Astronomy and Astrophysics, University of Chicago, 5640 South Ellis Avenue, Chicago, IL, 60637, USA}
\email{jc@astro.uchicago.edu}

\author[0000-0002-5751-1392]{J.~Carron}
\affiliation{Istituto ricerche solari Aldo e Cele Dacc\`o (IRSOL), Faculty of Informatics, Universit\`a della Svizzera italiana, 6605 Locarno, Switzerland}
\affiliation{Universit\'e de Gen\`eve, D\'epartement de Physique Th\'eorique, 24 Quai Ansermet, CH-1211 Gen\`eve 4, Switzerland}
\email{to.jcarron@gmail.com}

\author{C.~L.~Chang}
\affiliation{High-Energy Physics Division, Argonne National Laboratory, 9700 South Cass Avenue, Lemont, IL, 60439, USA}
\affiliation{Kavli Institute for Cosmological Physics, University of Chicago, 5640 South Ellis Avenue, Chicago, IL, 60637, USA}
\affiliation{Department of Astronomy and Astrophysics, University of Chicago, 5640 South Ellis Avenue, Chicago, IL, 60637, USA}
\email{clchang@uchicago.edu}

\author[0000-0002-5397-9035]{P.~M.~Chichura}
\affiliation{Department of Physics, University of Chicago, 5640 South Ellis Avenue, Chicago, IL, 60637, USA}
\affiliation{Kavli Institute for Cosmological Physics, University of Chicago, 5640 South Ellis Avenue, Chicago, IL, 60637, USA}
\email{pchichura@uchicago.edu}

\author{A.~Chokshi}
\affiliation{Department of Astronomy and Astrophysics, University of Chicago, 5640 South Ellis Avenue, Chicago, IL, 60637, USA}
\email{aman.chokshi@mcgill.ca}

\author[0000-0002-3091-8790]{T.-L.~Chou}
\affiliation{Department of Astronomy and Astrophysics, University of Chicago, 5640 South Ellis Avenue, Chicago, IL, 60637, USA}
\affiliation{Kavli Institute for Cosmological Physics, University of Chicago, 5640 South Ellis Avenue, Chicago, IL, 60637, USA}
\affiliation{National Taiwan University, No. 1, Sec. 4, Roosevelt Road, Taipei 106319, Taiwan}
\email{tlchou@uchicago.edu}

\author[0000-0002-2707-1672]{A.~Coerver}
\affiliation{Department of Physics, University of California, Berkeley, CA, 94720, USA}
\email{acoerver@berkeley.edu}

\author[0000-0001-9000-5013]{T.~M.~Crawford}
\affiliation{Department of Astronomy and Astrophysics, University of Chicago, 5640 South Ellis Avenue, Chicago, IL, 60637, USA}
\affiliation{Kavli Institute for Cosmological Physics, University of Chicago, 5640 South Ellis Avenue, Chicago, IL, 60637, USA}
\email{tmcrawfo@uchicago.edu}

\author[0000-0002-3760-2086]{C.~Daley}
\affiliation{Universit\'e Paris-Saclay, Universit\'e Paris Cit\'e, CEA, CNRS, AIM, 91191, Gif-sur-Yvette, France}
\affiliation{Department of Astronomy, University of Illinois Urbana-Champaign, 1002 West Green Street, Urbana, IL, 61801, USA}
\email{cailmd2@illinois.edu}

\author[0000-0001-5105-9473]{T.~de~Haan}
\affiliation{High Energy Accelerator Research Organization (KEK), Tsukuba, Ibaraki 305-0801, Japan}
\email{tijmen.dehaan@gmail.com}

\author{K.~R.~Dibert}
\affiliation{Department of Astronomy and Astrophysics, University of Chicago, 5640 South Ellis Avenue, Chicago, IL, 60637, USA}
\affiliation{Kavli Institute for Cosmological Physics, University of Chicago, 5640 South Ellis Avenue, Chicago, IL, 60637, USA}
\email{krdibert@uchicago.edu}

\author[0000-0001-7166-6422]{M.~A.~Dobbs}
\affiliation{Department of Physics and McGill Space Institute, McGill University, 3600 Rue University, Montreal, Quebec H3A 2T8, Canada}
\affiliation{Canadian Institute for Advanced Research, CIFAR Program in Gravity and the Extreme Universe, Toronto, ON, M5G 1Z8, Canada}
\email{matt.dobbs@mcgill.ca}

\author{M.~Doohan}
\affiliation{School of Physics, University of Melbourne, Parkville, VIC 3010, Australia}
\email{mdoohan@student.unimelb.edu.au}

\author[0000-0002-9962-2058]{D.~Dutcher}
\affiliation{Joseph Henry Laboratories of Physics, Jadwin Hall, Princeton University, Princeton, NJ 08544, USA}
\email{ddutcher@uchicago.edu}

\author[0000-0003-0307-9984]{T.~Eftekhari}
\affiliation{Center for Interdisciplinary Exploration and Research in Astronomy, Northwestern University, 1800 Sherman Avenue, Evanston, IL, 60201, USA}
\email{teftekhari@northwestern.edu}

\author{C.~Feng}
\affiliation{Department of Astronomy, University of Science and Technology of China, Hefei 230026, China}
\affiliation{School of Astronomy and Space Science, University of Science and Technology of China, Hefei 230026}
\affiliation{Department of Physics, University of Illinois Urbana-Champaign, 1110 West Green Street, Urbana, IL, 61801, USA}
\email{changfeng@ustc.edu.cn}

\author[0000-0002-4928-8813]{K.~R.~Ferguson}
\affiliation{Department of Physics and Astronomy, University of California, Los Angeles, CA, 90095, USA}
\affiliation{Department of Physics and Astronomy, Michigan State University, East Lansing, MI 48824, USA}
\email{kferguson@physics.ucla.edu}

\author[0000-0002-7130-7099]{N.~C.~Ferree}
\affiliation{California Institute of Technology, 1200 East California Boulevard., Pasadena, CA, 91125, USA}
\affiliation{Kavli Institute for Particle Astrophysics and Cosmology, Stanford University, 452 Lomita Mall, Stanford, CA, 94305, USA}
\affiliation{Department of Physics, Stanford University, 382 Via Pueblo Mall, Stanford, CA, 94305, USA}
\email{nferree@stanford.edu}

\author{K.~Fichman}
\affiliation{Department of Physics, University of Chicago, 5640 South Ellis Avenue, Chicago, IL, 60637, USA}
\affiliation{Kavli Institute for Cosmological Physics, University of Chicago, 5640 South Ellis Avenue, Chicago, IL, 60637, USA}
\email{kfichman@uchicago.edu}

\author[0000-0002-7145-1824]{A.~Foster}
\affiliation{Joseph Henry Laboratories of Physics, Jadwin Hall, Princeton University, Princeton, NJ 08544, USA}
\email{axf295@case.edu}

\author{S.~Galli}
\affiliation{Sorbonne Universit\'e, CNRS, UMR 7095, Institut d'Astrophysique de Paris, 98 bis bd Arago, 75014 Paris, France}
\email{gallis@iap.fr}

\author{A.~E.~Gambrel}
\affiliation{Kavli Institute for Cosmological Physics, University of Chicago, 5640 South Ellis Avenue, Chicago, IL, 60637, USA}
\email{anne.gambrel@gmail.com}

\author{A.~K.~Gao}
\affiliation{Department of Physics, University of Illinois Urbana-Champaign, 1110 West Green Street, Urbana, IL, 61801, USA}
\email{akgao2@illinois.edu}

\author[0000-0002-3833-8133]{F.~Ge}
\affiliation{California Institute of Technology, 1200 East California Boulevard., Pasadena, CA, 91125, USA}
\affiliation{Kavli Institute for Particle Astrophysics and Cosmology, Stanford University, 452 Lomita Mall, Stanford, CA, 94305, USA}
\affiliation{Department of Physics, Stanford University, 382 Via Pueblo Mall, Stanford, CA, 94305, USA}
\affiliation{Department of Physics \& Astronomy, University of California, One Shields Avenue, Davis, CA 95616, USA}
\email{fge@ucdavis.edu}

\author[0000-0001-7593-3962]{F.~Guidi}
\affiliation{Department of Physics \& Astronomy, University of California, One Shields Avenue, Davis, CA 95616, USA}
\affiliation{Sorbonne Universit\'e, CNRS, UMR 7095, Institut d'Astrophysique de Paris, 98 bis bd Arago, 75014 Paris, France}
\email{federica.guidi@iap.fr}

\author[0000-0003-2606-9340]{N.~W.~Halverson}
\affiliation{CASA, Department of Astrophysical and Planetary Sciences, University of Colorado, Boulder, CO, 80309, USA }
\affiliation{Department of Physics, University of Colorado, Boulder, CO, 80309, USA}
\email{nils.halverson@colorado.edu}

\author[0000-0003-1880-2733]{E.~Hivon}
\affiliation{Sorbonne Universit\'e, CNRS, UMR 7095, Institut d'Astrophysique de Paris, 98 bis bd Arago, 75014 Paris, France}
\email{hivon@iap.fr}

\author[0000-0002-9017-3567]{A.~Y.~Q.~Ho}
\affiliation{Department of Astronomy, Cornell University, Ithaca, NY 14853, USA}
\email{annayqho@cornell.edu}

\author[0000-0002-0463-6394]{G.~P.~Holder}
\affiliation{Department of Physics, University of Illinois Urbana-Champaign, 1110 West Green Street, Urbana, IL, 61801, USA}
\email{gholder@illinois.edu}

\author[0000-0002-5744-6439]{W.~L.~Holzapfel}
\affiliation{Department of Physics, University of California, Berkeley, CA, 94720, USA}
\email{swlh@cosmology.berkeley.edu}

\author{J.~C.~Hood}
\affiliation{Kavli Institute for Cosmological Physics, University of Chicago, 5640 South Ellis Avenue, Chicago, IL, 60637, USA}
\email{hoodjc@uchicago.edu}

\author{A.~Hryciuk}
\affiliation{Department of Physics, University of Chicago, 5640 South Ellis Avenue, Chicago, IL, 60637, USA}
\affiliation{Kavli Institute for Cosmological Physics, University of Chicago, 5640 South Ellis Avenue, Chicago, IL, 60637, USA}
\email{hryciuk@uchicago.edu}

\author[0000-0003-3595-0359]{N.~Huang}
\affiliation{Department of Physics, University of California, Berkeley, CA, 94720, USA}
\email{nhuang@alumni.princeton.edu}

\author{T.~Jhaveri}
\affiliation{Department of Astronomy and Astrophysics, University of Chicago, 5640 South Ellis Avenue, Chicago, IL, 60637, USA}
\affiliation{Kavli Institute for Cosmological Physics, University of Chicago, 5640 South Ellis Avenue, Chicago, IL, 60637, USA}
\email{tanishaj@uchicago.edu}

\author{F.~K\'eruzor\'e}
\affiliation{High-Energy Physics Division, Argonne National Laboratory, 9700 South Cass Avenue, Lemont, IL, 60439, USA}
\email{fkeruzore@anl.gov}

\author[0000-0002-8388-4950]{A.~R.~Khalife}
\affiliation{Sorbonne Universit\'e, CNRS, UMR 7095, Institut d'Astrophysique de Paris, 98 bis bd Arago, 75014 Paris, France}
\email{ridakhal@iap.fr}

\author[0000-0001-8053-8334]{L.~Knox}
\affiliation{Department of Physics \& Astronomy, University of California, One Shields Avenue, Davis, CA 95616, USA}
\email{lknox@ucdavis.edu}

\author{K.~Kornoelje}
\affiliation{Department of Astronomy and Astrophysics, University of Chicago, 5640 South Ellis Avenue, Chicago, IL, 60637, USA}
\affiliation{Kavli Institute for Cosmological Physics, University of Chicago, 5640 South Ellis Avenue, Chicago, IL, 60637, USA}
\affiliation{High-Energy Physics Division, Argonne National Laboratory, 9700 South Cass Avenue, Lemont, IL, 60439, USA}
\email{knk@uchicago.edu}

\author[0000-0003-4895-5247]{C.-L.~Kuo}
\affiliation{Kavli Institute for Particle Astrophysics and Cosmology, Stanford University, 452 Lomita Mall, Stanford, CA, 94305, USA}
\affiliation{Department of Physics, Stanford University, 382 Via Pueblo Mall, Stanford, CA, 94305, USA}
\affiliation{SLAC National Accelerator Laboratory, 2575 Sand Hill Road, Menlo Park, CA, 94025, USA}
\email{clkuo@stanford.edu}

\author[0000-0003-1792-2338]{T.~Laskar}
\affiliation{Department of Physics \& Astronomy, University of Utah, Salt Lake City, UT, 84112, USA}
\email{tanmoy.laskar@utah.edu}

\author{K.~Levy}
\affiliation{School of Physics, University of Melbourne, Parkville, VIC 3010, Australia}
\email{kevin.levy@student.unimelb.edu.au}

\author[0000-0002-4820-1122]{Y.~Li}
\affiliation{Kavli Institute for Cosmological Physics, University of Chicago, 5640 South Ellis Avenue, Chicago, IL, 60637, USA}
\email{yunyangl@uchicago.edu}

\author[0000-0002-4747-4276]{A.~E.~Lowitz}
\affiliation{Kavli Institute for Cosmological Physics, University of Chicago, 5640 South Ellis Avenue, Chicago, IL, 60637, USA}
\email{lowitz@arizona.edu}

\author{C.~Lu}
\affiliation{Department of Physics, University of Illinois Urbana-Champaign, 1110 West Green Street, Urbana, IL, 61801, USA}
\email{chunyul3@illinois.edu}

\author[0009-0004-3143-1708]{G.~P.~Lynch}
\affiliation{Department of Physics \& Astronomy, University of California, One Shields Avenue, Davis, CA 95616, USA}
\email{gplynch@ucdavis.edu}

\author[0000-0003-0976-4755]{T.~J.~Maccarone}
\affiliation{Department of Physics \& Astronomy, Box 41051, Texas Tech University, Lubbock TX 79409-1051, USA}
\email{thomas.maccarone@ttu.edu}

\author[0000-0002-4617-9320]{A.~S.~Maniyar}
\affiliation{Kavli Institute for Particle Astrophysics and Cosmology, Stanford University, 452 Lomita Mall, Stanford, CA, 94305, USA}
\affiliation{Department of Physics, Stanford University, 382 Via Pueblo Mall, Stanford, CA, 94305, USA}
\affiliation{SLAC National Accelerator Laboratory, 2575 Sand Hill Road, Menlo Park, CA, 94025, USA}
\email{abhishek.maniyar@lapth.cnrs.fr}

\author[0009-0009-0719-9781]{E.~S.~Martsen}
\affiliation{Department of Astronomy and Astrophysics, University of Chicago, 5640 South Ellis Avenue, Chicago, IL, 60637, USA}
\affiliation{Kavli Institute for Cosmological Physics, University of Chicago, 5640 South Ellis Avenue, Chicago, IL, 60637, USA}
\email{emartsen@uchicago.edu}

\author[0000-0002-1372-2534]{F.~Menanteau}
\affiliation{Department of Astronomy, University of Illinois Urbana-Champaign, 1002 West Green Street, Urbana, IL, 61801, USA}
\affiliation{Center for AstroPhysical Surveys, National Center for Supercomputing Applications, Urbana, IL, 61801, USA}
\email{felipe@illinois.edu}

\author[0000-0001-7317-0551]{M.~Millea}
\affiliation{Department of Physics, University of California, Berkeley, CA, 94720, USA}
\email{mariusmillea@gmail.com}

\author{J.~Montgomery}
\affiliation{Department of Physics and McGill Space Institute, McGill University, 3600 Rue University, Montreal, Quebec H3A 2T8, Canada}
\email{joshua.j.montgomery@gmail.com}

\author{Y.~Nakato}
\affiliation{Department of Physics, Stanford University, 382 Via Pueblo Mall, Stanford, CA, 94305, USA}
\email{yukanaka@stanford.edu}

\author[0000-0002-1328-2065]{T.~Natoli}
\affiliation{Kavli Institute for Cosmological Physics, University of Chicago, 5640 South Ellis Avenue, Chicago, IL, 60637, USA}
\email{tnatoli2@gmail.com}

\author[0000-0002-0963-7310]{Y.~Omori}
\affiliation{Department of Astronomy and Astrophysics, University of Chicago, 5640 South Ellis Avenue, Chicago, IL, 60637, USA}
\affiliation{Kavli Institute for Cosmological Physics, University of Chicago, 5640 South Ellis Avenue, Chicago, IL, 60637, USA}
\affiliation{NSF-Simons AI Institute for the Sky (SkAI), 172 E. Chestnut St., Chicago, IL 60611, USA}
\email{yuuki.om@gmail.com}

\author[0000-0003-0170-5638]{A.~Ouellette}
\affiliation{Department of Physics, University of Illinois Urbana-Champaign, 1110 West Green Street, Urbana, IL, 61801, USA}
\email{aaronjo2@illinois.edu}

\author[0000-0002-6164-9861]{Z.~Pan}
\affiliation{High-Energy Physics Division, Argonne National Laboratory, 9700 South Cass Avenue, Lemont, IL, 60439, USA}
\affiliation{Kavli Institute for Cosmological Physics, University of Chicago, 5640 South Ellis Avenue, Chicago, IL, 60637, USA}
\affiliation{Department of Physics, University of Chicago, 5640 South Ellis Avenue, Chicago, IL, 60637, USA}
\email{panz@uchicago.edu}

\author{P.~Paschos}
\affiliation{Enrico Fermi Institute, University of Chicago, 5640 South Ellis Avenue, Chicago, IL, 60637, USA}
\email{paschos@uchicago.edu}

\author[0000-0001-7946-557X]{K.~A.~Phadke}
\affiliation{Department of Astronomy, University of Illinois Urbana-Champaign, 1002 West Green Street, Urbana, IL, 61801, USA}
\affiliation{Center for AstroPhysical Surveys, National Center for Supercomputing Applications, Urbana, IL, 61801, USA}
\affiliation{NSF-Simons AI Institute for the Sky (SkAI), 172 E. Chestnut St., Chicago, IL 60611, USA}
\email{kphadke2@illinois.edu}

\author{K.~Prabhu}
\affiliation{Department of Physics \& Astronomy, University of California, One Shields Avenue, Davis, CA 95616, USA}
\email{kprabhu@ucdavis.edu}

\author[0009-0002-2589-5501]{W.~Quan}
\affiliation{High-Energy Physics Division, Argonne National Laboratory, 9700 South Cass Avenue, Lemont, IL, 60439, USA}
\affiliation{Department of Physics, University of Chicago, 5640 South Ellis Avenue, Chicago, IL, 60637, USA}
\affiliation{Kavli Institute for Cosmological Physics, University of Chicago, 5640 South Ellis Avenue, Chicago, IL, 60637, USA}
\email{weiquan@uchicago.edu}

\author{M.~Rahimi}
\affiliation{School of Physics, University of Melbourne, Parkville, VIC 3010, Australia}
\email{mahsa.rahimi@unimelb.edu.au}

\author[0000-0003-3953-1776]{A.~Rahlin}
\affiliation{Department of Astronomy and Astrophysics, University of Chicago, 5640 South Ellis Avenue, Chicago, IL, 60637, USA}
\affiliation{Kavli Institute for Cosmological Physics, University of Chicago, 5640 South Ellis Avenue, Chicago, IL, 60637, USA}
\email{arahlin@uchicago.edu}

\author[0000-0003-2226-9169]{C.~L.~Reichardt}
\affiliation{School of Physics, University of Melbourne, Parkville, VIC 3010, Australia}
\email{clreichardt@gmail.com}

\author{M.~Rouble}
\affiliation{Department of Physics and McGill Space Institute, McGill University, 3600 Rue University, Montreal, Quebec H3A 2T8, Canada}
\email{maclean.rouble@mail.mcgill.ca}

\author[0000-0001-5875-4751]{J.~E.~Ruhl}
\affiliation{Department of Physics, Case Western Reserve University, Cleveland, OH, 44106, USA}
\email{ruhl@case.edu}

\author[0000-0001-5755-5865]{A.~C.~Silva~Oliveira}
\affiliation{California Institute of Technology, 1200 East California Boulevard., Pasadena, CA, 91125, USA}
\affiliation{Kavli Institute for Particle Astrophysics and Cosmology, Stanford University, 452 Lomita Mall, Stanford, CA, 94305, USA}
\affiliation{Department of Physics, Stanford University, 382 Via Pueblo Mall, Stanford, CA, 94305, USA}
\email{anaoliv@stanford.edu}

\author{A.~Simpson}
\affiliation{Department of Astronomy and Astrophysics, University of Chicago, 5640 South Ellis Avenue, Chicago, IL, 60637, USA}
\affiliation{Kavli Institute for Cosmological Physics, University of Chicago, 5640 South Ellis Avenue, Chicago, IL, 60637, USA}
\email{simpsa@uchicago.edu}

\author[0000-0001-6155-5315]{J.~A.~Sobrin}
\affiliation{Department of Physics, Villanova University, 800 E Lancaster Ave, Villanova, PA 19085, USA}
\affiliation{Fermi National Accelerator Laboratory, MS209, P.O. Box 500, Batavia, IL, 60510, USA}
\email{joshua.sobrin@villanova.edu}

\author[0000-0002-5313-2782]{A.~A.~Stark}
\affiliation{Center for Astrophysics \textbar{} Harvard \& Smithsonian, 60 Garden Street, Cambridge, MA, 02138, USA}
\email{astark@cfa.harvard.edu}

\author{J.~Stephen}
\affiliation{Enrico Fermi Institute, University of Chicago, 5640 South Ellis Avenue, Chicago, IL, 60637, USA}
\email{jlstephen@uchicago.edu}

\author[0000-0002-6829-4389]{Q.~Taylor}
\affiliation{Department of Physics, Case Western Reserve University, Cleveland, OH, 44106, USA}
\affiliation{CERCA/ISO, Department of Physics, Case Western Reserve University, Cleveland, OH 44106, USA}
\email{qxt42@case.edu}

\author{C.~Trendafilova}
\affiliation{CERCA/ISO, Department of Physics, Case Western Reserve University, Cleveland, OH 44106, USA}
\affiliation{Center for AstroPhysical Surveys, National Center for Supercomputing Applications, Urbana, IL, 61801, USA}
\email{ctrendaf@illinois.edu}

\author[0000-0001-7192-3871]{J.~D.~Vieira}
\affiliation{Department of Astronomy, University of Illinois Urbana-Champaign, 1002 West Green Street, Urbana, IL, 61801, USA}
\affiliation{Department of Physics, University of Illinois Urbana-Champaign, 1110 West Green Street, Urbana, IL, 61801, USA}
\affiliation{Center for AstroPhysical Surveys, National Center for Supercomputing Applications, Urbana, IL, 61801, USA}
\email{jvieira@illinois.edu}

\author[0000-0002-4528-9886]{A.~G.~Vieregg}
\affiliation{Kavli Institute for Cosmological Physics, University of Chicago, 5640 South Ellis Avenue, Chicago, IL, 60637, USA}
\affiliation{Department of Astronomy and Astrophysics, University of Chicago, 5640 South Ellis Avenue, Chicago, IL, 60637, USA}
\affiliation{Enrico Fermi Institute, University of Chicago, 5640 South Ellis Avenue, Chicago, IL, 60637, USA}
\affiliation{Department of Physics, University of Chicago, 5640 South Ellis Avenue, Chicago, IL, 60637, USA}
\email{avieregg@kicp.uchicago.edu}

\author[0009-0009-3168-092X]{A.~Vitrier}
\affiliation{Sorbonne Universit\'e, CNRS, UMR 7095, Institut d'Astrophysique de Paris, 98 bis bd Arago, 75014 Paris, France}
\email{aline.vitrier@iap.fr}

\author{Y.~Wan}
\affiliation{Department of Astronomy, University of Illinois Urbana-Champaign, 1002 West Green Street, Urbana, IL, 61801, USA}
\affiliation{Center for AstroPhysical Surveys, National Center for Supercomputing Applications, Urbana, IL, 61801, USA}
\email{yujiew2@illinois.edu}

\author[0000-0002-3157-0407]{N.~Whitehorn}
\affiliation{Department of Physics and Astronomy, Michigan State University, East Lansing, MI 48824, USA}
\email{nathanw@msu.edu}

\author[0000-0001-5411-6920]{W.~L.~K.~Wu}
\affiliation{California Institute of Technology, 1200 East California Boulevard., Pasadena, CA, 91125, USA}
\affiliation{Kavli Institute for Particle Astrophysics and Cosmology, Stanford University, 452 Lomita Mall, Stanford, CA, 94305, USA}
\affiliation{SLAC National Accelerator Laboratory, 2575 Sand Hill Road, Menlo Park, CA, 94025, USA}
\email{kimwuu@gmail.com}

\author{M.~R.~Young}
\affiliation{Fermi National Accelerator Laboratory, MS209, P.O. Box 500, Batavia, IL, 60510, USA}
\affiliation{Kavli Institute for Cosmological Physics, University of Chicago, 5640 South Ellis Avenue, Chicago, IL, 60637, USA}
\email{mattyoungofficial@gmail.com}

\author[0000-0003-2375-0229]{J.~A.~Zebrowski}
\affiliation{Kavli Institute for Cosmological Physics, University of Chicago, 5640 South Ellis Avenue, Chicago, IL, 60637, USA}
\affiliation{Department of Astronomy and Astrophysics, University of Chicago, 5640 South Ellis Avenue, Chicago, IL, 60637, USA}
\affiliation{Fermi National Accelerator Laboratory, MS209, P.O. Box 500, Batavia, IL, 60510, USA}
\email{j.z@uchicago.edu}

\correspondingauthor{Sam~Guns}
\email{sguns@berkeley.edu}

\begin{abstract}
Wide-field instruments designed to observe the cosmic microwave background (CMB) have reached sufficient sensitivities and cadences to conduct untargeted millimeter-wave searches for astronomical transients. 
To date, untargeted searches have primarily discovered millimeter-wave stellar flares, while detections of extragalactic transients remain limited to a few flaring blazars and active galactic nuclei. In contrast, recent simulations of extragalactic transient populations suggest that current and future CMB experiments should be able to detect anywhere from a few to dozens of extragalactic transients per year from a range of sources (short- and long-duration gamma-ray bursts, tidal disruption events, fast blue optical transients, and others). At timescales shorter than one week, the predicted detections are dominated by flares from reverse shocks in long-duration gamma-ray bursts (GRBs), with some benchmark models predicting rates of $\sim10$ detections per year for the SPT-3G camera on the South Pole Telescope. In this work, we conduct a transient search optimized for the bright and brief millimeter-wave emission from GRB reverse shocks.
We update previous simulations by incorporating a more realistic model of the SPT-3G sensitivity, observing cadence, and detection algorithm, resulting in
nearly a factor of two improvement in detection sensitivity compared to prior estimates.
We subsequently carry out a search over four seasons of the $1500\,\rm{deg}^2$ SPT-3G Main field observations from 2019--2022, finding no candidate extragalactic events. We use the lack of detections to constrain the parameters of the synthesized GRB populations, setting upper limits on the fraction of GRBs with reverse shock components, $f_{\text{RS}}$, and the all-sky rate of reverse shocks beamed at Earth, $N_{\mathrm{all-sky}}$. For a standard $10^{51}\,\mathrm{erg}$ GRB with Lorentz factor $\Gamma=100$, we find limits of $f_{\text{RS}} < 0.57$ and $N_\mathrm{all-sky} < 1226$ bursts/year (at 90\% confidence). The most readily detectable model we consider is a %
$10^{52}\,\mathrm{erg}$
``dirty'' GRB with $\Gamma=50$, whose population we constrain to $f_{\text{RS}} < 0.09$ and $N_{\mathrm{all-sky}} < 120$ bursts/year.

\end{abstract}

\section{Introduction} \label{sec:intro}
\setcounter{footnote}{0}
In recent years, instruments designed to observe the cosmic microwave background (CMB) have been increasingly used to conduct searches for transient sources at millimeter wavelengths.
These efforts fill a critical gap in time-domain astronomy, as the recent proliferation of dedicated wide-field surveys has largely bypassed (sub-) millimeter wavebands.
For time-domain astronomy, untriggered wide-field searches provide a powerful complement to targeted follow-up observations with instruments like ALMA. They enable the discovery of new sources and new source populations, can track newly discovered sources over longer timescales, and are better suited for fast transients that occur on timescales shorter than the typical target-of-opportunity delay. 

The South Pole Telescope, with its 10-meter primary mirror and arcminute-scale resolution, was designed to observe the CMB and find millimeter-bright point sources and galaxy clusters. Data from the second-generation SPTpol camera were used to carry out the first ever untargeted search for millimeter-wave transients, finding one candidate event of uncertain origin in one year of observations of a 100 square degree field at \SI{150}{GHz} \citep{whitehorn16}. In recent years, untriggered transient detections by CMB observatories have become commonplace, starting with the serendipitous discovery of three bright stellar flares by the Atacama Cosmology Telescope (ACT) \citep{naess21} and the first results from a dedicated transient search using data from the third-generation SPT-3G camera on the South Pole Telescope \citep{guns21}. The SPT-3G search discovered 13 bright flares from nearby stars, as well as two candidate extragalactic events which were classified as likely active galactic nuclei (AGN) using optical follow-up observations. Both SPT and ACT have since carried out more transient searches, finding large numbers of stellar flares in the Galaxy and setting flux upper limits on millimeter emission from known extragalactic transients \citep{li23b, hervias24, tandoi24, biermann24}.

The fact that the vast majority of transient events discovered in SPT and ACT data consist of local stellar flares stands in contrast to theoretical efforts, which have mostly estimated millimeter-wavelength discovery rates for extragalactic transients.
Before the current generation of large-aperture CMB receivers came online, \cite{metzger15} predicted event rates for various classes of extragalactic transients such as on- and off-axis gamma-ray bursts (GRBs), tidal disruption events (TDEs), and binary neutron-star mergers, finding that a hypothetical CMB survey should expect to see on the order of 30 on-axis long-duration GRBs over its lifetime, but no significant number of any other extragalactic transient class. \cite{eftekhari22} expanded on these predictions by including both the forward and reverse shock emission from GRBs in their simulation suite, and by incorporating realistic survey areas and noise levels for various CMB experiments including SPT-3G. They confirmed the expectation that extragalactic event rates would be primarily due to on-axis GRB emission, with forward shocks dominant at long timescales ($\geq 1$ week) and reverse shocks at timescales shorter than 1 day. For a continuous 5.5-year SPT-3G-like survey, they predicted up to 40 reverse shock and 30 forward shock detections depending on the exact GRB model used. 

A GRB reverse shock (RS), in the context of the standard fireball model of GRBs \citep{meszaros93, meszaros97}, occurs when a secondary shock wave  travels through the GRB ejecta back towards the progenitor (hence moving in reverse in the frame of the primary, ``forward'' shock (FS) front formed by the ejecta colliding with the circumburst medium). The RS component has its own synchrotron spectrum that evolves independently from the forward shock spectrum, but the main spectral parameters scale with those of the forward shock spectrum according to
\begin{align}
S_{\nu,\mathrm{max}}^{\mathrm{RS}} &\sim \Gamma S_{\nu,\mathrm{max}}^{\mathrm{FS}}, \\
\nu_m^{\mathrm{RS}} &\sim \Gamma^{-2}\nu_m^{\mathrm{FS}}, 
\end{align}
where $\Gamma$ is the initial Lorentz factor of the GRB ejecta, and $S_{\nu,\mathrm{max}}^{\mathrm{RS}}$, $\nu_m^{\mathrm{RS}}$, $S_{\nu,\mathrm{max}}^{\mathrm{FS}}$, and $\nu_m^{\mathrm{FS}}$ are the peak flux and characteristic frequency of the RS and FS spectra, respectively \citep{sari99a}. For typical values of the shock Lorentz factor $\Gamma \sim 100$, the RS peaks in the millimeter-to-IR wavelength range and evolves on much faster timescales than the FS. 
Observing the RS component offers a powerful probe of GRB physics; in particular, it allows a direct measurement of  $\Gamma$, a crucial shock parameter which is not directly constrained by observations of forward shock emission. However, follow-up millimeter RS observations are technically difficult due to both the paucity of millimeter-wave observatories and the need to target the source within hours of discovery. Actual millimeter-wave observations of reverse shocks are rare, with only a limited number of follow-up detections to date \citep[e.g.][]{laskar13,perley14,laskar16,laskar18,laskar19a,laskar19b,dichiara22,keating26}. Still, the small sample of millimeter RS detections bears out the essential characteristics of reverse shocks relative to the forward shock component, and confirms their importance in gaining a full understanding of the early stages of GRBs across the entire electromagnetic spectrum. Directly finding GRBs in an untriggered millimeter-wave search would provide a powerful complement to all-sky monitoring capabilities at shorter wavelengths. Untriggered searches at optical wavelengths have already begun providing insight into poorly understood regions of GRB parameter space such as low-Lorentz factor ``dirty fireballs'' that are challenging to detect at optical wavelengths \citep{ho22}. 

The simulated reverse shock lightcurves in \cite{eftekhari22} have peak \SI{100}{GHz} fluxes ranging from tens to thousands of mJy, and durations from 20 minutes to 1 day. The predicted brightnesses and durations are similar to those of millimeter-wave flares from nearby stars, which we routinely detect with SPT-3G  \citep{tandoi24}. By contrast, the longer timescale of GRB forward shock emission necessitates different search strategies (stacking exposures over many days) and involves different backgrounds. The relevant background is expected to consist of long-duration AGN flares whose variability statistics are poorly understood at millimeter wavelengths. In this work, we therefore focus exclusively on the reverse shock emission and design a transient search optimized for short-duration signals.

In order to refine theoretical expectations, we adopt the RS population simulations from \cite{eftekhari22} but replace their simplified telescope observing model with an SPT-3G-specific forward simulation. The simplified observing model assumed full-field observations at a steady 12-hour cadence in a single \SI{100}{GHz} band with fixed \SI{6}{mJy} noise levels and a $10\,\sigma$ threshold for detection. The actual SPT-3G instrument observes the sky in three bands centered at \SI{95}{GHz}, \SI{150}{GHz}, and \SI{220}{GHz}, with time-varying noise levels and a cadence and scan strategy that involve many different timescales. To properly account for these effects, we  construct a detailed observing model which takes into account the SPT-3G observing cadence, detector coverage, scan strategy, and transient-finding method. In particular, the transient detection method we developed in \cite{whitehorn16} and \cite{guns21} optimally combines observations at different timescales and fully leverages the multi-band nature of SPT-3G data, and is therefore significantly more sensitive than a single-band $10\,\sigma$ threshold search. We feed the simulated RS lightcurves into our model and contrast the updated rate predictions with results from an untargeted search for short-duration transients in four years of SPT-3G data, using our search results to constrain the relative rate and peak millimeter wavelength luminosity of GRBs with bright reverse shock emission. 

\section{Methods} \label{sec:methods}
In this section, we first briefly describe the SPT-3G instrument and observation cadence. We then detail the GRB reverse shock simulations and describe a detection method that is optimized both for specific GRB reverse shock signatures and for bright, short-duration transients more generally.

\subsection{Instrument}
The South Pole Telescope is a 10-meter millimeter-wave telescope located at Amundsen-Scott South Pole Station, Antarctica \citep{carlstrom11}. The SPT Collaboration has used the SPT to survey the southern sky with three generations of cameras since 2007, producing maps of the CMB with an unprecedented combination of sensitivity and resolution. These maps have been used to produce CMB power spectra and associated cosmological parameter constraints \citep{dutcher21, balkenhol23, pan23, ge25, camphuis25}, as well as catalogs of galaxy clusters \citep{kornoelje25, bleem26} and millimeter-wave transients \citep{guns21, tandoi24}. The current SPT camera, SPT-3G, is the third-generation instrument installed on the SPT and has been operational since 2018. The SPT-3G focal plane contains $\sim 16,000$ transition edge sensor bolometers, split evenly between two linear polarizations and three observing bands centered at \SI{95}{GHz}, \SI{150}{GHz}, and \SI{220}{GHz}. The SPT's large primary mirror enables comparatively high-resolution observations for a mm-wave instrument, with diffraction-limited beams of $1.6\arcmin$, $1.2\arcmin$, and $1.0\arcmin$~FWHM at \SI{95}GHz, \SI{150}GHz, and \SI{220}GHz, respectively. A full description of the SPT-3G instrument design and performance can be found in \cite{sobrin22}. 

\subsection{Survey}
SPT-3G observes year-round, with the bulk of observing time dedicated to the SPT-3G Main survey which runs from late March through late November each year. The results presented here are based entirely on data from the Main survey. The SPT-3G Main  observing field comprises approximately 1500 square degrees of the Southern sky, bounded by lines of constant 
right ascension (RA) at $20^{\mathrm{h}}40^{\mathrm{m}}0^{\mathrm{s}}$ and $3^{\mathrm{h}}20^{\mathrm{m}}0^{\mathrm{s}}$,
and by lines of constant declination (dec) at $-70\arcdeg$ and $-42\arcdeg$.
The Main survey field is split into four subfields, each \SI{7.5}{\degree} tall in dec and spanning the entire width of the field. An SPT-3G observation is defined as a sequence of rastered scans covering a single subfield for a period of 2.2 hours. Over the course of an observation, the telescope scans back and forth at a constant $1\,$deg/s in azimuth across the subfield, before taking a \SI{12.5}{arcmin} upward step in elevation and repeating the process until the entire subfield has been covered. Because the focal plane is approximately 1.5 degrees tall and 1.5 degrees wide, a point source on the sky is observed for roughly 2 seconds each in 12 consecutive scans over the course of 20 minutes. 

\subsection{Mapmaking}
The time ordered data (TOD) for each detector is recorded and processed into sky maps for each observing frequency. 
We subtract a 3-year average from each map in order to remove backgrounds such as CMB, galaxy clusters, and static point sources.\footnote{We subtract a 3-year rather than 4-year average because the analysis was started before the conclusion of the 2022 observing season. This negligibly increases the final map noise.}
The difference maps are then filtered in order to remove large-scale atmospheric noise and maximize sensitivity to point sources. We then apply a noise estimation procedure and cut a small percentage of maps in which the filtered TOD noise is not sufficiently Gaussian. The mapmaking, filtering, noise estimation, and map cuts are described in detail in Appendix \ref{sec:app_maps}. 

\begin{figure}[t]
    \includegraphics[width=0.95\linewidth]{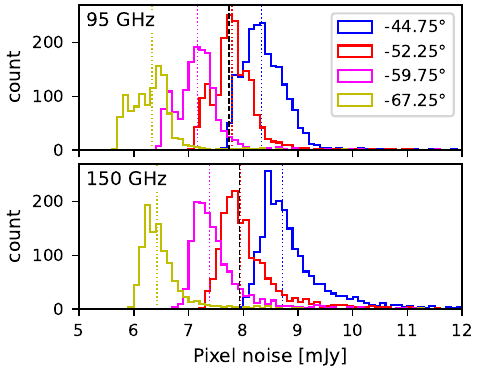}
    \caption{\label{fig:noise_histo} Median pixel noise for all observations in the 2019-2022 SPT-3G dataset, for the \SI{95}{GHz} (top) and \SI{150}{GHz} (bottom) observing bands. The colors denote the four survey subfields and are labeled with their center declination. The dotted and dashed lines show the median noise value for the observations in each subfield separately, and for all observations, respectively. The difference in noise values between subfields is primarily due to the relative sizes of the subfields and the fact that all fields are observed at the same azimuthal scan speed, with the highest declination subfield containing half the area of the lowest. Figure adapted from \cite{tandoi24}. }
\end{figure}

The observations used in this work span four observing seasons from 2019--2022, for a total of 6540 subfield maps across 1002 days of observing. The filtered single-observation maps in our dataset are dominated by white noise and have median RMS noise levels of 6.5--8.7 mJy depending on observing band and subfield, as shown in Figure \ref{fig:noise_histo}. We only use the \SI{95}{GHz} and \SI{150}{GHz} intensity maps in this work, since the \SI{220}{GHz} maps have substantially higher noise levels ($\ge$\SI{25}{mJy} RMS) and do not contribute meaningfully to a transient search for flat-spectrum sources. %

\subsection{Simulations}
\subsubsection{GRB Lightcurves}
\label{sec:rscurves}
\begin{figure*}%
\centering
    \includegraphics[width=0.95\textwidth]{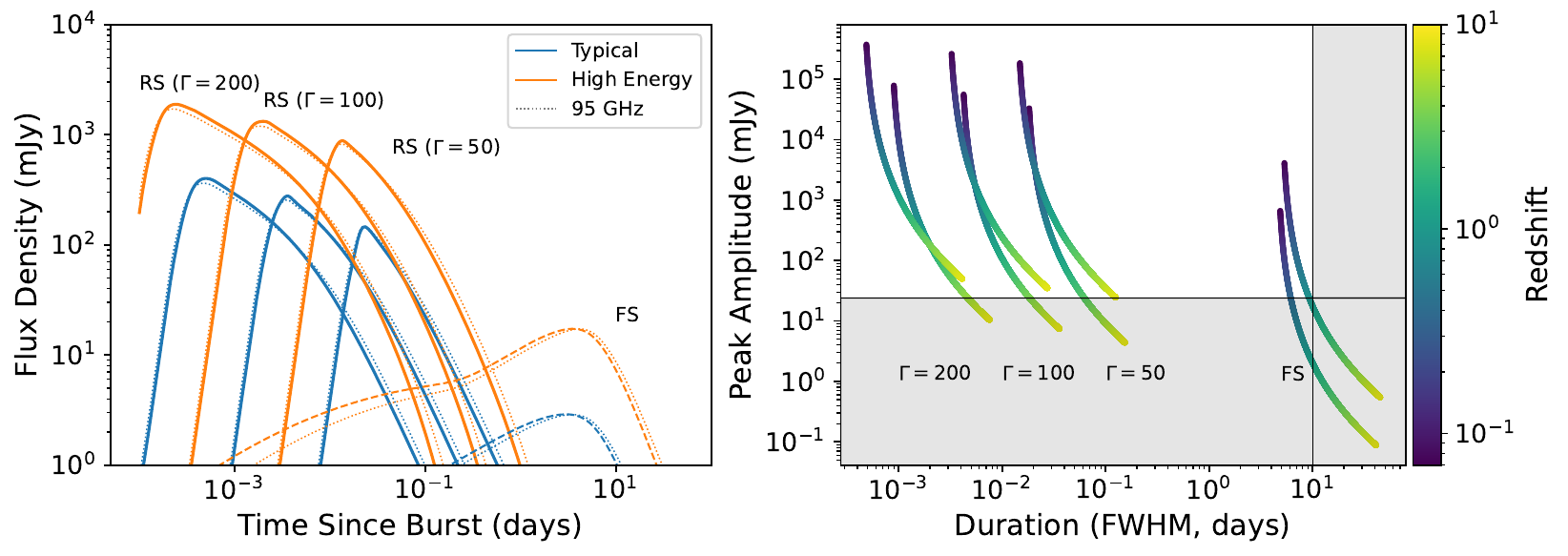}
    \caption{\label{fig:rscurves} GRB Forward Shock (FS) and Reverse Shock (RS) lightcurves adapted from \cite{eftekhari22}. \textbf{Left:} \SI{150}{GHz} lightcurves for all six simulated models shown at a reference redshift of $z=1$. The models are divided into two categories:  ``Typical'' bursts (blue) have kinetic energy $E_\mathrm{K} = 10^{51}\,\mathrm{erg}$ and density $n = 1\, \mathrm{cm}^{-3}$. Curves labeled ``High Energy'' (orange) correspond to burst kinetic energies of $E_\mathrm{K} = 10^{52}\,\mathrm{erg}$ and densities $n = 10\, \mathrm{cm}^{-3}$. For each category, we show the forward shock model (dashed) and three possible reverse shock models (solid) with initial Lorentz factors of $\Gamma=50$, $\Gamma=100$, or $\Gamma=200$. The corresponding \SI{95}{GHz} lightcurves are very similar to the \SI{150}{GHz} lightcurves and are shown with dotted lines. The spectral index across our observing bands is nearly flat. \textbf{Right:} Lightcurve peak amplitude and duration (FWHM) as a function of redshift for both the typical and high energy burst models. The simulated lightcurves span several orders of magnitude in flux and duration. The gray area approximately shows the effect of data cuts on observed flux ($>3\sigma$, with a median value of \SI{24}{mJy}) and duration ($< 10$ days FWHM). The true peak flux may be lower than observed for very short ($\mathrm{FWHM} < 10^{-2}$ days) flares. The brightest RS emission can in principle be detected out to $z=8$, while detections of FS emission are limited to $z\leq1$ even for the high energy model.}
\end{figure*}

Figure \ref{fig:rscurves} shows the reverse shock lightcurves used by \cite{eftekhari22} for their simulation suite.
They considered six different parameter choices for the modeled reverse-shock (RS) signals,  
originating from on-axis top-hat jets 
with two choices for the overall luminosity and three for the initial Lorentz factor $\Gamma$.\footnote{Only two values of $\Gamma$ (50 and 200) were used in \cite{eftekhari22}, but the authors also produced lightcurves for $\Gamma = 100$ which we incorporate into our simulations.}
The typical-luminosity bursts assume a burst kinetic energy $E_\mathrm{K} = 10^{51}\,$erg and an interstellar medium with a density of $n = 1\, \mathrm{cm}^{-3}$. The ``high energy''\footnote{We adopt the terminology of \cite{eftekhari22} and use ``high energy'' to denote bursts with high injected energy, not to be confused with bursts that produce high energy photons.}bursts have $E_\mathrm{K} = 10^{52}\,$ erg and $n = 10\, \mathrm{cm}^{-3}$. The corresponding isotropic gamma-ray luminosities are $L_{ \mathrm{iso}}=5\times10^{51}\,$erg s$^{-1}$ and $5\times10^{52}\,$erg s$^{-1}$ for the typical and high energy bursts, respectively. Each GRB is then assumed to produce a millimeter-bright reverse shock, evaluated with three possible initial Lorentz factors ($\Gamma=50, 100$, or $200$) for a total of six reverse shock models. All models assume Newtonian reverse shocks and fiducial values of $p = 2.5, \epsilon_e = 0.1$, and $\epsilon_B=0.01$ for the electron energy power law index, relative electron energy density, and relative magnetic field energy density, respectively. We summarize the simulation parameters and give the  specific luminosity at \SI{150}{GHz} for each RS model in Table \ref{table:rsparams}. 
The resulting lightcurves are shown in the left panel of Figure \ref{fig:rscurves} at a reference redshift of $z=1$, along with the corresponding forward shock lightcurves (which are independent of $\Gamma$). The reverse shock lightcurves are up to a factor of 100 brighter at peak than the forward shock emission, and evolve on timescales $10^2$--$10^4$ times faster. In order to populate a cosmological distribution of bursts, the lightcurves are scaled to different redshifts by applying cosmological time dilation to the lightcurves and scaling the fluxes with the corresponding luminosity distances. While the simulations include independent lightcurves at \SI{95}{GHz} and \SI{150}{GHz}, the lightcurves have nearly flat spectra in our bands and $k$-corrections are ignored when scaling the flux. 
The right panel of Figure \ref{fig:rscurves} shows the peak flux and duration of all models as a function of redshift. In particular, the duration of the RS lightcurves is less than the duration of a single 2.2-hour SPT-3G observation out to redshifts of $z=4$ for the $\Gamma = 50$ models and beyond $z=10$ for the fast-evolving $\Gamma=200$ models, suggesting they would be detected as excess flux in a single observation only. Both the low-redshift $\Gamma=50$ bursts and all $\Gamma=100$ and $200$ bursts show variations on timescales of one or several 100-second telescope scans. A burst that appears in an SPT-3G subfield during an observing window is not guaranteed to be observed, 
for example, if it fades before the telescope has scanned to the correct elevation. The fast time evolution necessitates simulating the RS signals at the timescale of single scans in order to accurately estimate the average flux captured by the SPT-3G detectors. The forward shock emission on the other hand evolves on much longer timescales of 4--40 days, with only the nearest bursts at $z<1$ lasting shorter than 10 days. 

\begin{table}[h!]
\hspace{-5em}
\renewcommand{\arraystretch}{1.6}
\begin{tabular}{|c|l|c|c|}
\cline{3-4}
 \multicolumn{2}{c|}{}  & \textbf{$\Gamma$} & $L_{\nu}^{150\,\mathrm{GHz}}\; (\mathrm{erg}\,\mathrm{s}^{-1}\mathrm{Hz}^{-1})$ \\ \hline
\multirow{3}{*}{\begin{sideways}\textbf{typical}\end{sideways}} & $E_\mathrm{K}=10^{51}\mathrm{erg}$ & 50 & $6.1\times 10^{32}$ \\ \cline{3-4} 
 &$n=1\,\mathrm{cm}^{-3}$ & 100 & $1.2\times 10^{33}$ \\ \cline{3-4} 
 &  $L_{\mathrm{iso}}=5\times10^{51}\mathrm{erg\,s}^{-1}$ & 200 & $1.8\times 10^{33}$ \\ \hline
\multirow{3}{*}{\begin{sideways}\textbf{high energy}\end{sideways}} & $E_\mathrm{K}=10^{52}\mathrm{erg}$ & 50 & $3.9\times 10^{33}$ \\ \cline{3-4} 
 & $n=10\,\mathrm{cm}^{-3}$ & 100 & $5.8\times 10^{33}$ \\ \cline{3-4} 
 & $L_{\mathrm{iso}}=5\times10^{52}\mathrm{erg\,s}^{-1}$ & 200 & $8.2\times 10^{33}$ \\ \hline
\end{tabular}
\caption{\label{table:rsparams}Input parameters for the different GRB reverse shocks used in our simulation pipeline. All shocks assume a top-hat jet expanding into a constant-density interstellar medium, with non-relativistic reverse shocks. The microphysical parameters are set to $p = 2.5$, $\epsilon_e = 0.1$, and $\epsilon_B=0.01$.}
\end{table}

\subsubsection{Population Synthesis}
\begin{figure}[t]
    \includegraphics[width=0.95\linewidth]{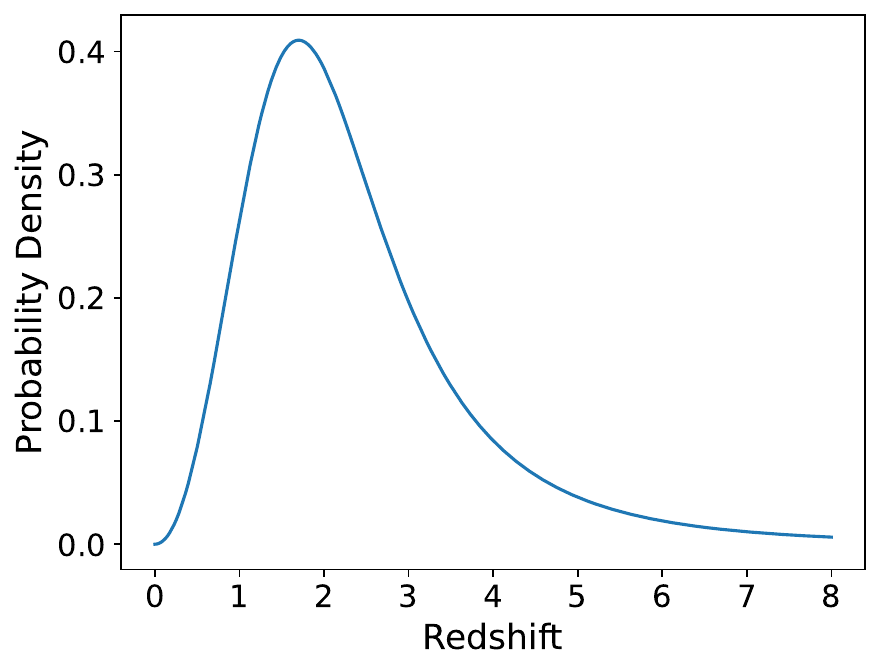}
    \caption{\label{fig:zhist} Redshift distribution of GRBs injected into our simulation pipeline. The distribution tracks the star formation rate of \cite{madau14} and is peaked at $z=1.7$.}
\end{figure}

In order to simulate a cosmological distribution of bursts, local volumetric rates ($\mathcal{R} = 0.2\;\mathrm{Gpc}^{-3} \mathrm{yr}^{-1}$ for the typical events, $\mathcal{R} = 0.1\;\mathrm{Gpc}^{-3} \mathrm{yr}^{-1}$ for the high-energy bursts, from \citealt{sun15}) were scaled with the star formation rate from \citet{madau14} 
\begin{equation}
\label{eq:sfr}
    \mathrm{SFR}(z) = 0.015 \frac{\left(1+z\right)^{2.7}}{1+\left[\left(1+z\right)/2.9\right]^{5.6}} M_{\sun}\,\mathrm{year}^{-1}\,\mathrm{Mpc}^{-3},
\end{equation}
and integrated over a standard universe with $H_0 = 67.74$ km $\mathrm{s}^{-1}\,\mathrm{Mpc}^{-1}$, $\Omega_m = 0.3089$, and $\Omega_{\Lambda} = 0.6911$ \citep{planck16-47} in order to create a sampling distribution between $z = 0$ and a maximal redshift $z_{\mathrm{max}}=10$. The resulting normalized distribution is shown in Figure \ref{fig:zhist} and is peaked around $z=1.7$. For each of the six burst models, we sample 40,000 bursts from the redshift distribution and inject the corresponding lightcurves (scaled and time-dilated) at uniformly sampled times and locations in four simulated observing seasons of SPT-3G data. For any given search algorithm, we convert the simulated detection efficiency $\eta$ of each class of bursts to a predicted number of detections $N_{\mathrm{sim}}$ using 
\begin{align}
     N_{\mathrm{sim}} &= \mathcal{A}\eta \Delta T\int_0^{z_{\mathrm{max}}}\frac{\mathrm{SFR}(z)}{\mathrm{SFR}(0)}\frac{\mathcal{R}}{1+z}\frac{dV_c(z)}{dz} dz,\\
         &= \label{eq:allsky}\frac{\mathcal{A}}{4\pi}\eta \Delta T N_{\mathrm{all-sky},}
\end{align}
where $\mathcal{A}$ is the survey area in steradians, $\eta$ the detected fraction of injected lightcurves, $\Delta T$ the total survey duration in years, $\mathcal{R}$ the volumetric rate, $dV_c(z)/dz$ the differential comoving volume, and $N_{\mathrm{all-sky}}$ the all-sky rate (in events/year).  
\subsubsection{SPT-3G Simulation Pipeline}
\cite{eftekhari22} used a simplified telescope model in which the entire SPT-3G Main field was observed for 15 minutes every 12 hours in a single frequency band centered at \SI{100}{GHz}, with no seasonal breaks. A simulated reverse shock was counted as observed if its peak flux during the observing window reached \SI{60}{mJy}, or $10\,\sigma$ assuming a \SI{6}{mJy} noise level. 
In order to get a more accurate estimate of the number of observed events, we construct a detailed transient observing model for the SPT-3G Main field that realistically accounts for the observing cadence and noise levels by using noise measurements and timestamps from actual SPT-3G observations. For each telescope scan in each observation, we record the instantaneous flux of all simulated reverse shocks as seen by the individual detectors. The signal is formed by multiplying the simulated flux with the SPT-3G beam pattern at each frequency, and weighting each pixel using the data-derived per-scan detector weights stored during the mapmaking steps outlined in Appendix \ref{sec:app_maps}. This creates a realistic thumbnail map of the observed point source signal at the telescope scan level, which is then averaged over the course of an observation and added to an unfiltered noisy patch of sky randomly sampled from a difference map of a real observation. The result is an accurate simulation of the flux, weight, and noise of a time-dependent point source integrated over a realistic SPT-3G observation. We filter the resulting thumbnail maps using the same filtering as the data pipeline, and recover the injected source using our standard transient finding method. 

\subsubsection{Generic Lightcurves}
Finally, we also use our simulation pipeline to inject simple Gaussian and exponential flares at a variety of peak flux levels and timescales. For these simulations, we only calculate detection efficiencies, without converting the efficiencies to on-sky rates. These efficiencies can be used to forecast the sensitivity of SPT-3G transient searches to a wide variety of transient sources, or to measure or constrain their populations from search results. 

\subsection{Transient Detection}
\label{sec:transient_finding}
We design a transient search that is sensitive to the simulated reverse shock signal described in Section \ref{sec:rscurves}, but also broadly to any transient signal on timescales of 10 days or less. 
The duration of 10 days was chosen specifically to be sensitive to rapidly evolving reverse shock components while isolating the search from the poorly understood long-term variability backgrounds of AGN.
This follows from an earlier SPT-3G transient search that found two distinct populations consisting of flares from nearby stars in the Galaxy on timescales of tens of minutes to several days, and flares from normally quiescent active galactic nuclei (AGN) at longer timescales \citep{guns21}. By focusing on short-duration transients only, we avoid having to identify and characterize potential signals from two distinct source populations. The long-duration flares from \cite{guns21} had multiple possible counterparts in the IR and optical, and required dedicated follow-up observations to be classified as AGN, whereas the short-duration stellar flares are all clearly associated with bright, nearby stars in Gaia DR3 \citep{gaia16,gaia23} and furthermore are predominantly M dwarfs and active binaries, classes of stars that are also strongly active at other wavelengths.

We use substantially the same transient finding algorithm as previous SPT transient searches \citep{whitehorn16,guns21,tandoi24}. The method, described in detail below, relies on constructing pixel lightcurves in fixed-width windows and fitting a parametric flare model with variable amplitude and duration to the data. Other millimeter transient searches \citep{li23b,biermann24} and simulations \citep{metzger15,eftekhari22} rely on finding outlier pixels either in single-observation maps or in map stacks (averages over a fixed duration window). The main advantage of our method is that all possible timescales can be probed simultaneously, as shown in \cite{guns21} where we found both single-observation stellar flares and multi-week AGN flares in the same search. Even for this search where we focus exclusively on short-duration flares, the range of possible timescales  spans from 2 hours to 10 days. By explicitly fitting the flare start time and duration, our algorithm effectively finds the optimal map stack that results in maximal signal-to-noise ratio for any given transient source. 

The main disadvantage is that running a nonlinear optimizer on a lightcurve for each map pixel is more computationally expensive than simply searching for outliers in single maps or fixed-duration map stacks. We reduce the number of map pixels to be searched by combining both methods, first imposing a preliminary significance cut directly on the maps and only applying the fitting method to pixels that pass the cut. We set the initial pixel cut to only include lightcurves that have a simultaneous $3\sigma$ fluctuation in both the $95$ and $150\,$GHz bands anywhere in the time window. As we show in Section \ref{sec:res_grbrs}, this cut sets a detection floor of approximately \SI{24}{mJy} on all transients, while having almost no impact on detecting the simulated reverse shock signals. 

\subsubsection{Detection Algorithm}
For lightcurves that pass the initial cut, we fit a model with four parameters: independent amplitudes in each band $S_b$, a common start time $t_0$, and flare duration parameter $w$. 
The log likelihood of the lightcurve data in each band $\phi_b(t)$ takes the form 
\begin{equation}
    -2\mathrm{ln}\mathcal{L}(\theta) = \sum_t \sum_{b\in\{95\mathrm{GHz},\atop 150\mathrm{GHz}\}}\frac{\left(\phi_b(t) - f_b(t, \theta)\right)^2}{\sigma_b^2(t)},
\end{equation}
where $t$ indexes the time of each observation and $f_b$ is a parametric flare profile depending on time and a vector of parameters $\theta$ = $(S_{95\mathrm{GHz}}, S_{150\mathrm{GHz}}, t_0, w)$.
The joint likelihood formulation assumes that the noise in the two bands is statistically independent. 
The filtering described in Appendix~\ref{sec:filtering} is designed to remove atmospheric fluctuations that could result in noise correlation between the bands. 
In the case where some residual correlated noise remains, the limits presented in this work will be  overestimates.
The fitting is done by constructing and maximizing a test statistic (TS) as a function of the parameters:
\begin{equation}
\label{eq:ts}
    \text{TS} = 2\ln L(\hat{\theta}) - 2\ln L(0) + \ln \left( \frac{{\hat{w}}}{\Delta T} \right),
\end{equation}
where hatted quantities denote best-fit estimates and $\Delta T = 12$ days is the lightcurve window size. 
The first two terms in Eq. \ref{eq:ts} are a log likelihood ratio of the best-fit model to the null model with zero amplitudes. 
The third term implements a logarithmic penalty in the fitted duration, which accounts for the larger trial phase space of shorter timescales \citep{braun10}. Without this penalty, random white-noise fluctuations on very short single-scan timescales would yield higher likelihood ratios than genuine longer-duration events, creating a baseline skew.

Prior SPT analyses only used a Gaussian profile for $f$,
\begin{equation}
    f_b(t, \theta) = S_b\,\mathrm{exp}\left[\frac{-\left(t-t_0\right)^2}{2 w^2}\right],
\end{equation}
intended as a generic catch-all model. Because we are specifically interested in transients with very fast or instantaneous rise times and exponential decays, we also fit a simple exponential profile:
\begin{equation}
    f_b(t, \theta) = S_b\,\mathrm{exp}\left[\frac{-\left(t-t_0\right)}{w}\right],
\end{equation}
for $t \geq t_0$ and $f_b(t,\theta) = 0$ for $t < t_0$. 
For each event, we compare the test statistics resulting from both the Gaussian and exponential fits and keep the higher of the two. For injected 12-hour exponential flares near the detection threshold, using the correct profile increases our sensitivity by $5\%$ on average. The increase is relatively modest because in the low SNR limit, only a few observations contribute meaningfully to detection, and differentiating between different flare profiles is difficult or impossible. The sensitivity increase is larger for brighter flares but also becomes less important for the purpose of detection, since even a large profile mismatch will still result in an above-threshold TS value. 

\subsection{Background Estimation}
\label{sec:threshold}
Following \cite{whitehorn16}, we use our data to measure the background of noise-induced events and set a detection threshold based on their TS distribution. In the single-band search in \cite{whitehorn16}, this was done by classifying the events with a negative fit amplitude as background events, whereas the positive amplitude events could come from either background or signal. As we use both \SI{95}{GHz} and \SI{150}{GHz} data in our search, we construct the background distribution from events that have negative amplitudes in both bands. We run the positive and negative searches separately and invert the initial pixel cut for the negative search, only evaluating lightcurves with simultaneous $<-3\sigma$ amplitude in both bands. Events with one positive and one negative amplitude are discarded in the search for sources in both the real data and in the simulations of injected sources. The negative and positive amplitude events are not fully independent: the TOD filter described in Appendix \ref{sec:app_maps} will create negative amplitude artifacts around bright positive sources. To remove these from the background dataset, we cut every negative event at the same declination and within half a degree in RA of any bright ($\mathrm{TS}>100$) positive event.

\section{Results} \label{sec:results}
\subsection{Data}
\label{sec:res_data}
\begin{figure*}[h!t]
    \centering
    {\includegraphics[width=0.95\linewidth]{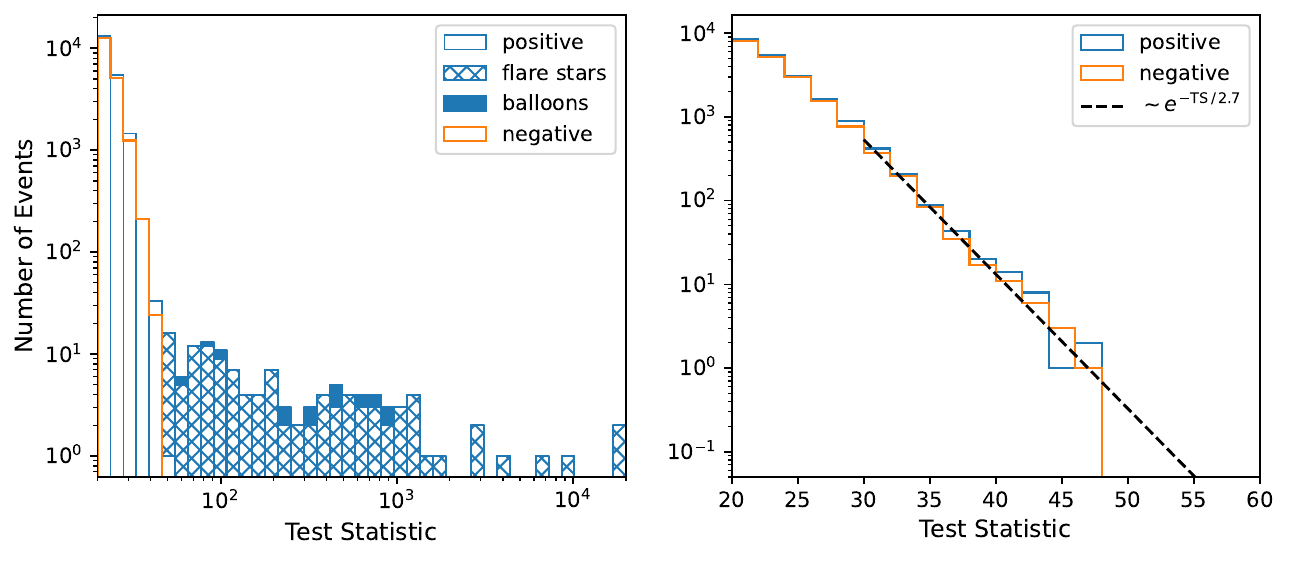}}
    \caption{\label{fig:tshisto} TS distribution for all same-sign events in the 4-year SPT-3G transient event dataset. Events are separated by the sign of their fit amplitudes at 95 GHz and 150 GHz: either both positive or both negative. The negative events are induced by map noise and are used to estimate the background rate.  
    \textbf{Left}: Final dataset of negative events (orange) and positive events (blue). Positive events associated with nearby stars are marked with a cross-hatch pattern, and those associated with weather balloons are shown in solid blue. 
    \textbf{Right}: Final event distribution with balloons and matches to nearby stars removed. The dashed line is the maximum-likelihood fit of an exponential distribution $\mathrm{d}n_{b}/\mathrm{dTS} \propto \mathrm{exp} \left(-\mathrm{TS}\middle/\tau\right)$ to the tail of the negative events, with $\tau = 2.7$. 
    }
\end{figure*}

We search the four-year SPT-3G Main dataset with the transient finding method described in Section \ref{sec:transient_finding} and keep any lightcurves that pass our data cuts and where both fit amplitudes are either positive or negative. The resulting TS histograms are shown in Figure \ref{fig:tshisto}. The negative (background) TS histogram shows an exponentially decaying tail that ends at TS = 46.2. The positive TS histogram has an additional 128 events with higher TS and extends out to $\mathrm{TS} = 18499$. We inspect these 128 events manually and categorize them as follows: 
\begin{itemize}
    \item \textbf{Known SPT-3G stellar flares:} We recover 110 out of the 111 flares described in the previously published SPT-3G 4-year stellar flare catalog \citep{tandoi24}. One event from the catalog is not recovered due to the stricter map cuts employed in this analysis (see Appendix \ref{sec:app_mapcuts}).
    \item \textbf{Weather balloons:} 
    11 events show strong evidence of originating from rapidly moving weather balloons launched daily by the South Pole Meteorological Office. In \cite{guns21}, we cut these events by demanding that any transient be visible in two or more subsequent observations. We have since established a data sharing agreement with the South Pole Meteorological Office, which allows us to track the weather balloon GPS transmitters and cut any telescope scan that intersects with a balloon GPS track. This system is successful more than $95$\% of the time. The only failure mode occurs when the GPS transmitter becomes separated from the balloon after launch (usually during the descent phase, after the balloon has ruptured in the upper atmosphere) and untracked debris moves through the telescope's field of view. Any balloon-sourced signals can be easily identified by their thermal spectral index, short-duration (one telescope scan), and movement-induced streaking. Because weather balloons are launched on a regular schedule, we also cross-check with the balloon launch logs. 
    \item \textbf{Previously unknown stellar flares:}
    We run a slightly updated version of the \textit{Gaia} DR3 cross-matching algorithm from \cite{tandoi24} on remaining events and find that five of them have likely matches to nearby stars. Four have association p-values (as defined in \citealt{tandoi24}) far below the threshold of $p=0.020$, and one is just above at $p=0.021$. One event is associated with a star already present in the SPT-3G catalog. These events were rejected from the SPT-3G flare star analysis (before any Gaia associations were made) because they failed a strict visual assessment. Upon closer examination, all five events appear consistent with being dim flares that are perturbed by nearby noise fluctuations. The stellar counterparts of the four new flaring stars are compatible with the wider population described in \cite{tandoi24}, and are discussed in more detail in Appendix \ref{sec:app_stars} along with a description of the updated cross-matching algorithm. 
    \item \textbf{Unclassified events:} The two remaining significant positive events are not obviously associated with any source (Galactic or otherwise) in Gaia and are detected at TS values of 46.3 and 47.8, respectively. We do not assign any particular classification to these low-significance events and keep them in the dataset. 
\end{itemize}

These classifications are shown in Figure \ref{fig:tshisto}. Removing all the events classified as stars or balloons leaves a distribution very close to that of the negative events, but with a small fractional excess of positive events.
After removing the classified events, the dataset of events with ${\rm TS}>20$ contains 20354 positive-amplitude and 19197 negative-amplitude events, or a $6\%$ excess in the positive events. 
If the distribution of events is modeled as a noise-dominated Gaussian, the observed excess can be explained by a systematic offset of only $0.012\sigma$ in the transient detection algorithm.
Such a small offset in the map zero point would have a completely negligible impact on the number of high-significance (${\rm TS}>48$) events and result in very slightly more conservative upper limits on detected sources.
Regardless of the presence of this low-TS excess, we find that after removing identifiable flares (Galactic and man-made) both the positive and negative distributions contain no events beyond $\mathrm{TS}=48$, and no candidate extragalactic transients remain in the dataset above this value. 

For the purpose of background estimation we fit an exponential distribution to the tail of the negative event histogram. As discussed in Appendix \ref{sec:app_maps}, we expect the first few bins of the TS distribution to be incomplete, which can be confirmed visually in Figure \ref{fig:tshisto}. Limiting the fit to $\mathrm{TS}>30$ gives a differential background density of 
\begin{equation}
\label{eq:bgrate}
    \frac{d n_{b}}{d\mathrm{TS}} = \frac{723}{2.7}\;\mathrm{exp}\left(-\frac{\mathrm{TS}-30}{2.7}\right),
\end{equation}
where 723 is the number of negative events at $\mathrm{TS}\geq 30$ and $2.7$ is the maximum likelihood estimate of the exponential decay rate. For a given detection threshold we can then integrate this density to obtain an estimate of the background count. 

\subsection{Simulations}
\label{sec:res_grbrs}
\begin{figure}
\centering
    \includegraphics[width=0.95\linewidth]{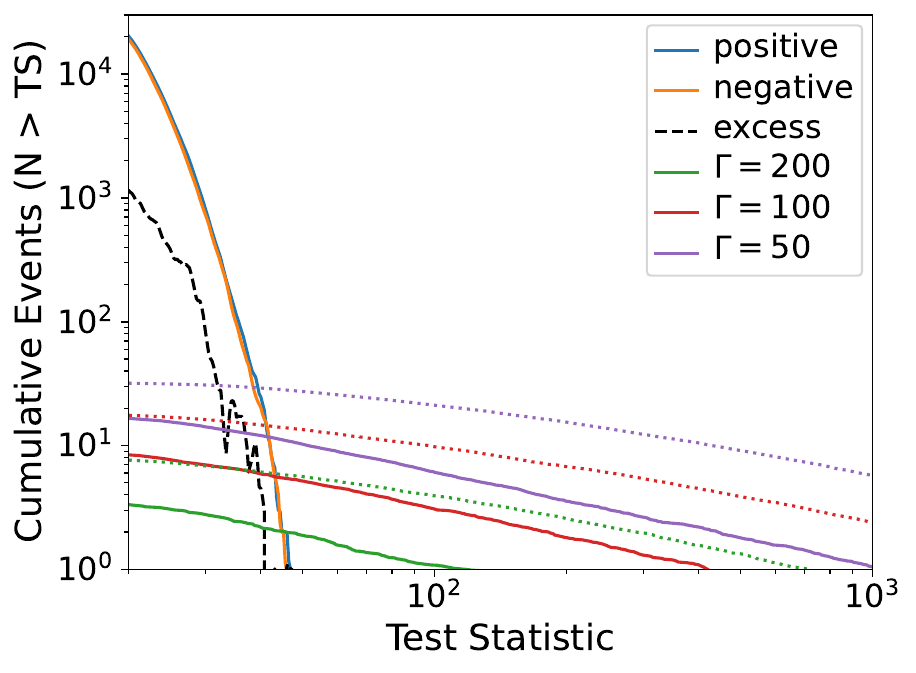}
    \caption{\label{fig:rssims} Cumulative number counts for the positive (blue) and negative (orange) data distributions and the simulated number counts for all six reverse shock models. The models are color-coded by Lorentz factor (green for $\Gamma = 200$, red for $\Gamma = 100$, and purple for $\Gamma = 50$). The typical shocks are shown with solid lines and the high energy shocks with dotted lines. We also show the difference between the positive and negative counts with the black dashed line. This fraction of  excess positive events is consistent with a negligible skewness in the noise distribution or tiny  ($0.012\sigma$) offset in the zero point of maps.} %
\end{figure}
The cumulative TS counts of the simulated reverse shock populations are shown in Figure \ref{fig:rssims}, along with the background (negative amplitude) and signal (positive amplitude) data distributions, where we have removed Galactic and man-made sources brighter than the brightest background event from the latter. In contrast with the two data-derived distributions, which terminate by TS=48, all six simulated reverse shock populations slowly decay out to very high TS values. 
Every simulated distribution skews toward higher significance, producing a larger population of high-TS events relative to low-TS events.
For example, the $\Gamma = 50$ high energy population contains a total of $27.7$ events at TS $> 48$, compared to $4.5$ events at $20 < \mathrm{TS} < 48$. 
The choice of $\mathrm{TS}=48$ as a threshold is somewhat arbitrary; in practice, different thresholds are used for different purposes. If the goal is to discover new source populations, the threshold should be set to the lowest possible value that does not overly strain targeted follow-up resources with false positives from background events. Using the integrated version of Eq. \ref{eq:bgrate}, $\mathrm{TS}=48$ corresponds to a low false positive rate of $\sim1$ event in four years of data. For the purpose of setting limits on theoretical models, a more conservative (higher) threshold leads to better background separation and tighter limits when the uncertainty in the background rate is properly included. In Section \ref{sec:constraints} we use thresholds of $\mathrm{TS}=53-57$ for the statistical analysis of the GRB reverse shock models. We keep $\mathrm{TS}=48$ as a general purpose ``detection threshold'' both for setting limits on generic (non-GRB) reference models (Section \ref{sec:refmodels}), and in order to allow a comparison with the single-band detection metric from \cite{eftekhari22}, where a GRB is considered detected if its \SI{100}{GHz} flux\footnote{While 100 GHz is slightly higher than our nearest observing frequency of 95 GHz, the simulated GRBs have nearly flat spectral indices ($-0.1 < \alpha < 0.25$) and the error from the frequency difference is at most 1\%.} is above $10\,\sigma$ in a single observation. We note that the underlying noise measurement has both temporal and spatial variations (see Figures \ref{fig:noise_histo} and \ref{fig:thumbnails}), so $10\,\sigma$ can range from approximately $60-100$ mJy. 

As shown in Table \ref{table:rssims}, when compared to a single-band $10\,\sigma$ search, the two-band algorithm used in this search detects up to 2.7 times more reverse shock events depending on the RS model used. All RS models have relatively flat spectral indices in our bands, so the increase in sensitivity can be easily explained by adding a second observing band with similar noise and by noting that
the test statistic asymptotically follows a  $\chi^2$ distribution with 2 degrees of freedom. In the limit where the data perfectly match the model, this reduces to \(\text{TS} \sim (\text{signal}/\text{noise})^2\), meaning that $\mathrm{TS} = 48$ corresponds heuristically to a signal with flux of $\sim7\sigma$ in one band or $\sim5\sigma$ in two bands. These simulations include the efficiency of the data cuts (pixel significance and fit FWHM) described in Section \ref{sec:transient_finding}. Above the $\mathrm{TS}=48$ threshold, the cuts have a negligible effect and reduce the number of detectable sources by $\leq 1\%$.

\section{GRB Model Constraints}
The simulated mono-energetic GRB RS populations shown in Figure~\ref{fig:rssims} and Table~\ref{table:rssims} predict a relatively flat TS distribution with a tail of high-significance events, and the absence of high-TS events constrains the parameters of these models.
However, the true underlying GRB RS population is not mono-energetic, and one might ask whether a more realistic GRB RS population, with a higher abundance at lower energies, could reproduce the observed excess of positive events.
This interpretation is disfavored by the very steep slope of the excess positive event distribution.
If we model the flux distribution of GRB RS events as a power law, $N(>S_\nu) \propto S_{\nu}^{\gamma}$, the power law scaling of the GRB RS flux distribution would need to be $\gamma < -20$ to explain the measured excess and still be consistent with the lack of high-TS events observed.
This is far steeper than any measured or modeled distribution of mm-wavelength GRB FS or RS fluxes which are generally shallower than the Euclidean prediction ($\gamma>-1.5$) in the low source flux regime \citep{metzger15,eftekhari22}.
Therefore, we conclude that no reasonable population of GRB RS events can significantly contribute to the observed excess of low TS positive events without producing a population of high-TS events. 
As was discussed in Section~\ref{sec:res_data}, the observed slight excess of positive events is consistent with a negligible skewness in the noise distribution or a tiny ($0.012\sigma$) offset in the detection algorithm.
We proceed to use the lack of high significance (${\rm TS}>48$) events to constrain GRB RS models.

While the simulated reverse shock lightcurves and underlying populations depend on many parameter choices, we choose to focus on two parameters in particular. The simulations assume that every on-axis GRB produces a millimeter-bright reverse shock. The true fraction of reverse-shock-producing GRBs is poorly measured. Targeted millimeter-wave follow-up observations suggest a lower limit of $\geq 30\%$ \citep{eftekhari22}. We introduce a scaling parameter $f_{\mathrm{RS}}$ representing the fraction of GRBs with a reverse shock and use our data to constrain it independently for each model, which can be done easily since the number of detected events above some fixed threshold scales linearly with $f_{\mathrm{RS}}$. 
The assumed local volumetric rate of the progenitor population is degenerate with the fraction of bursts that produce a bright reverse shock component; consequently, the upper limits on $f_{\mathrm{RS}}$ presented here are conditional upon the underlying baseline volumetric rates ($\mathcal{R}$) and star formation history adopted in our population synthesis framework.
The second parameter is the millimeter-band specific luminosity $L_{\nu}^{\mathrm{mm}}$. In principle, this is a function of the energy, circumburst density, and microphysical parameters of the simulated GRB, all of which are fixed at fiducial values. However, it is conceptually far simpler to set upper limits on the observable peak flux directly and leave open to theoretical input the question of how to alter the underlying models in order to conform to the observations. 

\subsection{GRB RS Parameter Constraints}
\label{sec:constraints}
The most straightforward constraints on the parameters of interest can be obtained by integrating the data and simulation distributions above a given TS threshold into a set of cumulative detection counts, obtaining an upper limit from the observed absence of data events, and converting the data upper limit into limits on the simulation parameters of interest. We convert a given TS threshold into a cumulative background rate, $n_b$, by integrating Eq. \ref{eq:bgrate}, and use Feldman--Cousins confidence regions \citep{feldman98} in order to calculate 90\% signal upper limits $N_{\alpha}(N,n_b)$, where $\alpha = 0.9$, from the expected number of background events $n_b$ and the observed number of signal events, $N$, above the threshold.
In all cases we consider, the most restrictive constraints come from TS thresholds where the cumulative signal counts $N=0$. For values of TS threshold where $N=0$ and $n_b \rightarrow 0$, $N_\alpha (0,0)$ asymptotes to  $2.44$ for $\alpha=0.9$. In all cases, the TS threshold that results in the most restrictive model constraints has corresponding background counts $n_b<1$. 

Scaling the simulated counts by a model-dependent factor $A$ we then derive a corresponding upper limit 
\begin{equation}
    A_{\alpha}(N=0, n_b) = \frac{N_{\alpha}(0,n_b)}{N_{\mathrm{sim}}},
\end{equation}
where $N_{\mathrm{sim}}$ denotes the number of simulated events above the detection threshold. Limits on the parameters of interest can then be derived easily. The fraction of GRBs with reverse shocks, $f_{\mathrm{RS}}$, can be directly identified with the scaling parameter $A$, which represents the fraction by which the expected number counts need to be reduced to match the observed data. In order to constrain the millimeter luminosity independently of $f_{\mathrm{RS}}$, we calculate $N_{\mathrm{sim}}$ for each of the reduced-luminosity simulations and use linear interpolation between the simulation points to map the constraint on $A$ into a constraint on $L^{\mathrm{mm}}_{\nu}$ for each model. 

\begin{figure*}
\centering
    \includegraphics[width=0.95\textwidth]{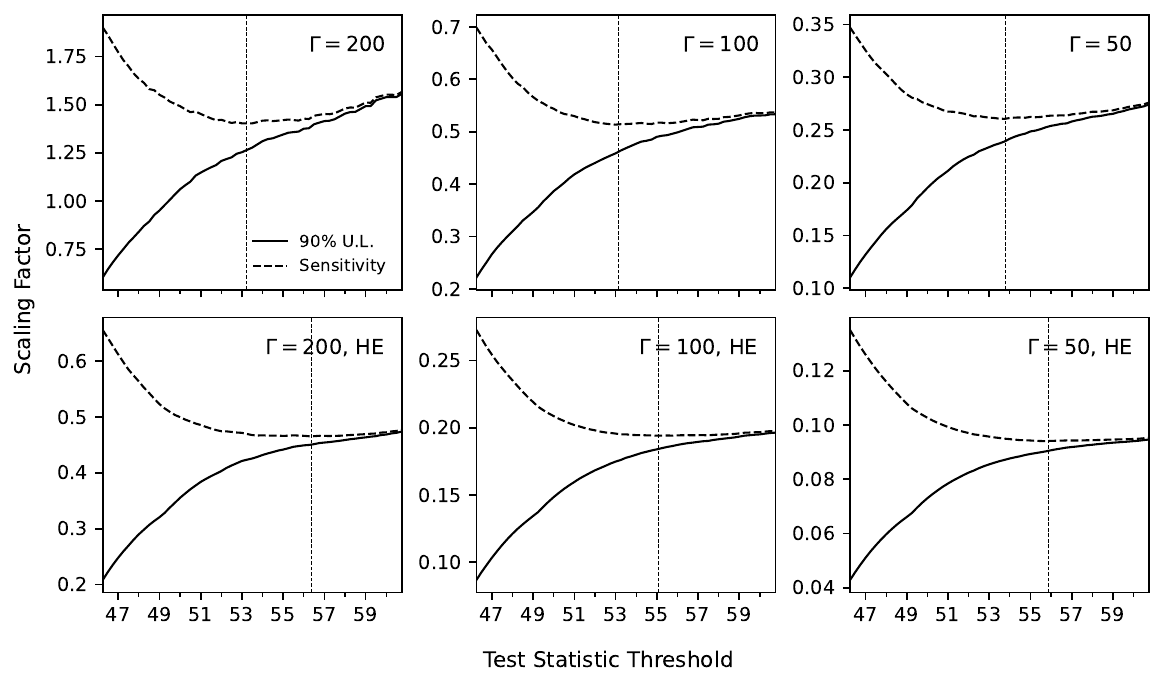}
    \caption{\label{fig:sensi} 
    The six panels show constraints on six different GRB RS models. In each panel, the solid line shows the $90\%$ confidence region upper limit on the model scaling parameter resulting from no observed detections and the expected instrumental background rate determined from the observed distribution of negative counts. The dashed curve shows the sensitivity of the upper limit on the TS threshold. The TS threshold for evaluating the model upper limits is taken as the value  which minimizes the sensitivity. The $90\%$ confidence region upper limit on the scaling parameter for each model is given by the solid curve evaluated at the minimum sensitivity TS threshold value marked by the dotted vertical line.}
\end{figure*}

The resulting model constraints vary as a function of the chosen TS threshold. Using the lowest TS value with no counts as the threshold is not an unbiased choice and has the potential to underestimate the true upper limits on parameters. In order to determine the TS threshold in an unbiased way, we use the minimum sensitivity method of \cite{hill02}. The sensitivity, $\bar{A_{\alpha}}$, of the scaling parameter $A_{\alpha}$ is defined as the average expected upper limit (at $\alpha$ confidence) on $A$ for a background $n_b$,
\begin{equation}
    \bar{A}_{\alpha}(n_b) = \sum_{k=0}^{\infty}\frac{n_b^k\,e^{-n_b}}{k!} A_{\alpha}\left(k, n_b\right).
    \label{eqn:minsens}
\end{equation}
This is a weighted sum over upper limits on the scaling parameter $A_{\alpha}(k,n_b)$, where each term represents the upper limit from detecting $k$ counts with $n_b$ expected events, weighted by the corresponding Poisson probability.  
The threshold is then set by choosing the value of $n_b$ (and the corresponding TS threshold) that minimizes $\bar{A}_{\alpha}(n_b)$ for the desired upper limit confidence region, $\alpha$. 
The $90\%$ confidence region upper limits are given by the value of $A_{\alpha}(N=0, n_b)$ where $n_b$ is the number of expected background counts at the ``minimum sensitivity" TS threshold.
Figure \ref{fig:sensi} shows both $A_{\alpha}(N,n_b)$ and $\bar{A}_{\alpha}(n_b)$ for $\alpha=0.9$ as a function of the TS threshold for all six GRB models.
The two quantities diverge at lower TS thresholds because, while the upper limit from a detection of zero counts decreases, the lower probability of observing zero counts reduces the contribution of these terms to the sum in Eq. \ref{eqn:minsens}. The  models considered produce a range of ``minimum sensitivity" TS thresholds between $\mathrm{TS} = 53$ and $\mathrm{TS} = 56$, with the minimum value marked by a dotted vertical line. The value of the threshold and resulting $f_{\mathrm{RS}}$ constraints for each model are given in Table \ref{table:rsul}. At the fiducial luminosity, the resulting upper limit on $f_{\mathrm{RS}}$ ranges from 0.09 for the $\Gamma = 50$ high energy model with the highest predicted event rate, to unconstrained for the typical \(\Gamma = 200\) model (where the unscaled simulations remain fully compatible with the data at the $90\%$ confidence level).

\begin{figure}[h]
\centering
    \includegraphics[width=0.95\linewidth]{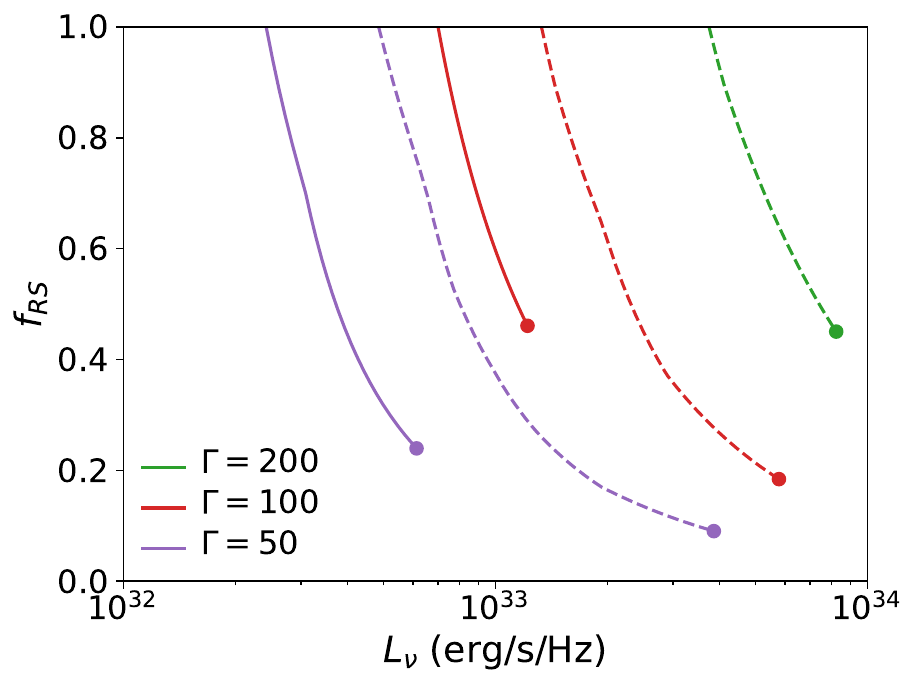}
    \caption{\label{fig:contours} Combined 90\% upper limits on the fraction $f_{\mathrm{RS}}$ of GRBs with a reverse shock, and the specific luminosity $L_{\nu}$ at $150\,$GHz. Colors are the same as in Figure \ref{fig:rssims}. Solid lines denote typical energy models and dashed lines their high energy equivalent. The fiducial luminosity of each model is shown with a circular marker. }
\end{figure}

We show the combined constraints on $f_{\mathrm{RS}}$ and $L_{\nu}^\mathrm{mm}$ in Figure \ref{fig:contours}. We do not simulate any luminosities that are higher than the fiducial luminosity for each model (marked with a solid circle). Again taking the $\Gamma = 50$ high energy model that would produce the most detections as an example, fixing $f_{\mathrm{RS}}$ to the literature lower limit of 0.3 constrains the \SI{150}{GHz} luminosity to be $\leq 1.2\times 10^{33}$~erg~s$^{-1}$~Hz$^{-1}$, a factor of three lower than the fiducial value. Where applicable, similar constraints on $L_{\nu}^\mathrm{mm}$ for fixed values of $f_{\mathrm{RS}} = 1.0$ and $f_{\mathrm{RS}} = 0.3$ are given in Table \ref{table:rsul}. 

Finally, as an alternative and simpler approach, we fix the luminosities to their fiducial values and set direct constraints on the all-sky rate of Eq. \ref{eq:allsky}. The upper limits are given by 
\begin{equation}
    N_{\mathrm{all-sky}} \leq \frac{4\pi N_{\alpha}}{\eta\mathcal{A}\Delta T},
\end{equation}
which only depends on the (model- and threshold-dependent) detection efficiency $\eta$, and the survey area $\mathcal{A}$ and duration $\Delta T$. By constraining $N_{\mathrm{all-sky}}$ directly, this approach sidesteps the issue of whether the excess in simulation counts is due to an overestimate of the rate of reverse shocks $f_{\mathrm{RS}}$, the volumetric rate of GRBs $\mathcal{R}$, or any other degenerate parameter that linearly scales $N_{\mathrm{all-sky}}$. We show the detection efficiencies $\eta$ and 90\% upper limits on $N_{\mathrm{all-sky}}$ in Table \ref{table:rsul}; the limits range from $\leq 120$ events per year (all-sky) for the high-energy $\Gamma = 50$ model, where the combination of high flux and long duration ensures that we detect almost 20\% of all injected flares, to $\leq$ 3332 events per year for the $\Gamma = 200$ model, whose extremely short-duration flares (two minute FWHM at $z=1$) are observable by SPT-3G less than 1\% of the time. 

\begin{table}
\centering
\begin{tabular*}{\columnwidth}{l@{\extracolsep{\fill}}rr}
\toprule 
    \multicolumn{1}{l}{Model} &  
    \multicolumn{1}{r}{$\mathrm{TS}>48$} &
    \multicolumn{1}{r}{$10\,\sigma\,(95\,\mathrm{GHz})$} \\
    \midrule
    $\Gamma = 50$ &  10.6 & 3.8 \\%
    $\Gamma = 100$ &  5.3 & 2.0 \\%
    $\Gamma = 200$ & 2.9 & 0.7 \\%
    $\Gamma = 50$, HE & 27.7 & 15.0 \\%
    $\Gamma = 100$, HE & 13.7 & 6.9 \\%
    $\Gamma = 200$, HE & 5.7 & 2.7 \\%
    \bottomrule
\end{tabular*}
\caption{\label{table:rssims} Predicted detection counts for the reverse shock simulations. Detections are calculated using either a $\mathrm{TS} > 48$ threshold or a $10\,\sigma$ detection in the \SI{95}{GHz} band only.}
\end{table}

\begin{table*}
\centering
\begin{tabular*}{\textwidth}{l@{\extracolsep{\fill}}rr|rrrr|}
    \toprule
    \toprule
    \multicolumn{1}{c}{} & 
    \multicolumn{2}{c}{} &
    \multicolumn{4}{c}{90\% Upper Limits} \\
    \multicolumn{1}{l}{Model} &  
    \multicolumn{1}{r}{$\mathrm{TS}_{\mathrm{sens}}$} &
    \multicolumn{1}{r}{$\eta (\%)$} &
    \multicolumn{1}{r}{$f_{\mathrm{RS}}$} &
    \multicolumn{1}{r}{$N_{\mathrm{all-sky}}$} &
    \multicolumn{1}{r}{$L_{\nu}/L_{\nu}^f (f_{\mathrm{RS}} = 1)$} &   
    \multicolumn{1}{r}{$L_{\nu}/L_{\nu}^f (f_{\mathrm{RS}} = 0.3)$} \\  
    \midrule
    $\Gamma = 50$ &  53.8 & 3.2 & 0.24 & 634 & 0.39 & 0.85 \\
    $\Gamma = 100$ &  53.1 & 1.6 & 0.57 & 1226 & 0.57 & $>1.0$ \\
    $\Gamma = 200$ & 53.2 & 0.6 & $>1.0$ & 3332 & $>1.0$ & $>1.0$ \\
    $\Gamma = 50$, HE & 55.9 & 17.3 & 0.09 & 120 & 0.13 & 0.31 \\
    $\Gamma = 100$, HE & 55.1 & 8.4 & 0.18 & 245 & 0.23 & 0.62 \\
    $\Gamma = 200$, HE & 56.4 & 3.5 & 0.45 & 600 & 0.46 & $>1.0$ \\
    \bottomrule
\end{tabular*}
\caption{\label{table:rsul}Detection thresholds, efficiencies, and model upper limits for the reverse shock simulations. $\bm{\mathrm{TS}_{\mathrm{sens}}}$: the model-dependent detection threshold as described in Section \ref{sec:constraints}. $\bm{\eta}$: percentage of injected events that pass data cuts and are detected at $\mathrm{TS} > \mathrm{TS}_{\mathrm{sens}}$. Upper limits are given on $\bm{f_{\mathrm{RS}}}$: the fraction of GRBs with reverse shock emission; $\bm{N_{\mathrm{all-sky}}}$: the all-sky rate (in events per year); and $\bm{L_{\nu} / L_{\nu}^f}$: the specific luminosity (at \SI{150}{GHz}), expressed as a fraction of the fiducial luminosity in Table \ref{table:rsparams}.}
\end{table*}
\begin{figure*}
    \centering
    \includegraphics[width=0.48\textwidth]{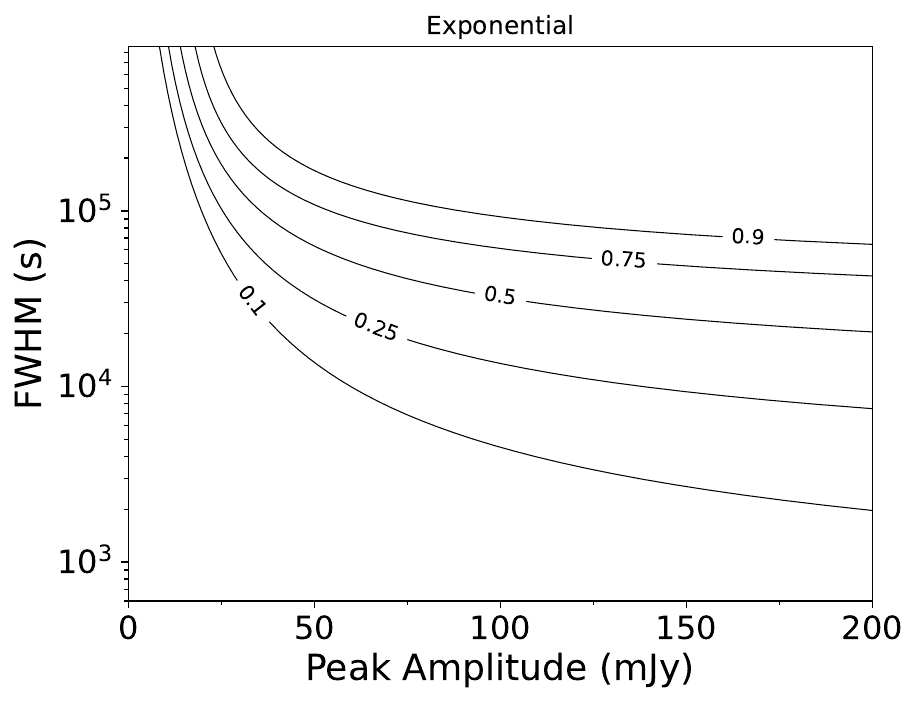}
    \hfill
    \includegraphics[width=0.48\textwidth]{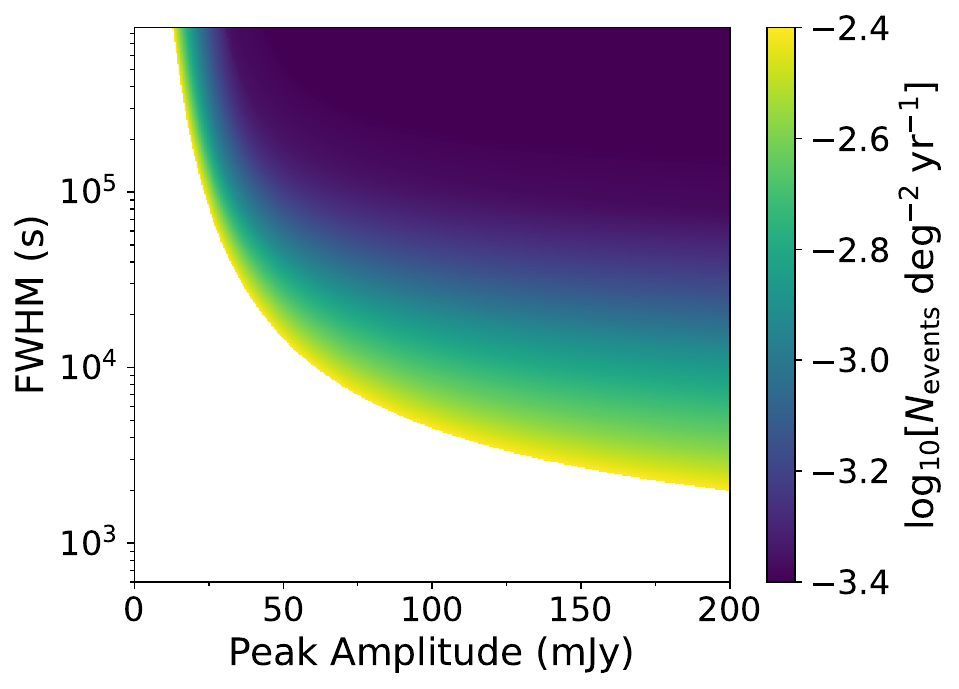}
    \caption{\label{fig:exp_nev} \textbf{Left:} Detection efficiency for injected exponential flares. \textbf{Right:} 90\% rate upper limits (in number of events per square degree per year) for exponential flares with timescales ranging from 10 minutes to 10 days FWHM and peak fluxes from 5 to 200 mJy. The data upper limit is calculated using a TS=48 threshold, and the simulations use the same cuts and threshold. }
\end{figure*}

 \subsection{Reference Models}
 \label{sec:refmodels}
In addition to number counts for the GRB simulations, we also use our observing model to calculate overall detection efficiencies for flat-spectrum, instant-rise, exponential decay flares. We inject 10,000 flares for each combination of flux and duration, covering flares with peak fluxes from 5 to $200\,$mJy and durations from 10 minutes to 10 days FWHM. We calculate the detection efficiency (defined as the fraction of flares that pass the cuts and exceed the $\mathrm{TS}=48$ detection threshold) and convert it into a 90\% upper limit on the sky rate (in events per square degree per year) using the same approach we used for the reverse shock simulations in Section \ref{sec:constraints}. The detection efficiencies for the exponential flares and the resulting rate limits are shown in Figure \ref{fig:exp_nev}. In general, our detection efficiency at fixed duration rapidly drops off below fluxes of 30-60 mJy, and at higher fluxes plateaus to fixed values largely determined by the interplay of flare duration and observing cadence. At fixed peak flux above the detection threshold, longer flares are detected with increasing efficiency up until the regime of very long and bright ($>100$ mJy peak, $>1$ day duration) flares that are detected nearly 100\% of the time. When we repeat the simulation with symmetric Gaussian flares, the resulting efficiencies differ by less than 20\% from the exponential case across the sampled parameter space. As noted in \cite{tandoi24}, the SPT-3G cadence is not in general fast or uniform enough to differentiate between different models of time evolution, even for very bright stellar flares. The efficiencies and limits shown here should allow one to calculate SPT-3G detection sensitivities to good approximation for any postulated class of astrophysical transients with single peaks of a given brightness and duration.

\section{Conclusion} \label{sec:conclusion}
We have carried out a search for short-duration extragalactic transients in four seasons of SPT-3G Main survey data, covering 1500 square degrees of the southern sky for approximately 1000 days of observing with an average cadence of 12 hours. Our transient detection method maximizes a modified likelihood ratio (test statistic or TS) under the assumption of a simple flare model (Gaussian or exponential) with four parameters: peak amplitudes in two SPT observing bands centered at \SI{95}{GHz} and \SI{150}{GHz}, a peak time, and flare duration (FWHM). Using events with negative amplitudes as an estimate of the noise background, we derive a detection cutoff of TS = 48. The TS of a detected flare is a function of both flare amplitude and duration and our observing cadence and noise. The detection limit of TS = 48 corresponds roughly to a 50\% detection probability for a 1-day 35 mJy flare, 8-hour 100 mJy flare, or 4-hour 200 mJy flare. We provide a full set of Gaussian and exponential simulations to map flare parameters to SPT-3G detection probabilities.

The main result of this paper can be stated simply: no significant short-duration extragalactic signals exist in our dataset. Of the 128 positive events with TS higher than the most significant negative event, 11 are sourced by weather balloons that separated from their GPS unit and could not be tracked by our software.
An additional 115 are associated with nearby stars, 110 of which were published in the 2019--2022 SPT-3G flare star catalog \citep{tandoi24}. 
The two remaining positive events are not obviously associated with any source (Galactic or otherwise), but are detected at significance below our detection threshold ($\rm{TS}=46.3$ and $47.8$) and are consistent with noise fluctuations.  

We compare the null result of our search with the results of a suite of simulations of millimeter-wave lightcurves of reverse shock emission from gamma-ray bursts. These simulations adapt and extend the framework of \cite{eftekhari22}, using their lightcurves as input to our multi-band transient finding method. We find that our method outperforms a simple single-band, single-observation significance search by a factor of $\sim 2$, with the improvement increasing with flare duration. The simulations consist of reverse shock emission at two energy scales corresponding to GRB kinetic energies $E_\mathrm{K} = 10^{51}\,$erg and $E_\mathrm{K} = 10^{52}\,$erg,  and three different peak Lorentz factors of $\Gamma=50, 100, \mathrm{and}\, 200$ for a total of six models. Nearly all simulated models predict a number of detections several times higher than the observed SPT-3G upper limits.
We find that either lowering the fraction of GRBs with reverse shock emission by factors of 2--10 or fixing it to the reference lower limit of 30\% while scaling down the burst luminosity by factors of 1.2--3 resolves the discrepancy and brings the simulations in line with the lack of detections. 

\cite{eftekhari22} predicted GRB RS detection rates for both SPT-3G and ACT. The number of predicted detections for the 7-year ACT survey is approximately four times lower than for a 5.5-year survey with SPT-3G. The ACT collaboration carried out a systematic search for extragalactic transients and detected no events above $5\,\sigma$ in five years of observations \citep{biermann24}. Their results are consistent with the lack of detections reported in this work. 

The results presented here represent the first systematic constraints on GRB reverse-shock population parameters from millimeter wavelength observations, enabled by realistic simulations of detector sensitivities and noise properties, and telescope observing strategy and cadence. In the future, SPT-3G will release the results of transient searches from observing periods and survey fields that were not considered in this work, and from searches extended to longer-duration bursts (including the GRB forward shock component). For the latter, the main background signal is expected to be caused by variability in dim active galactic nuclei \citep{guns21}, a population which is poorly understood at millimeter wavelengths and is the subject of active investigation. 

The transient finding method and simulation pipeline developed in this and prior SPT-3G transient searches can easily be adapted to different observing cadences and noise levels, including those of new CMB experiments such as Advanced Simons Observatory (ASO) and the planned SPT-3G+ receiver that will search for transients with significantly increased mapping speed \citep{abitbol25,athreya26}. The stellar flare catalogs and GRB model limits produced by SPT-3G can be used to inform design choices and search strategies in order to maximize future millimeter-wave transient science output.

\section*{Acknowledgments} \label{sec:ack}
The South Pole Telescope program is supported by the National Science Foundation (NSF) through awards OPP-1852617 and OPP-2332483. Partial support is also provided by the Kavli Institute of Cosmological Physics at the University of Chicago. 
Argonne National Laboratory’s work was supported by the U.S. Department of Energy, Office of High Energy Physics, under contract DE-AC02-06CH11357. 
The UC Davis group acknowledges support from Michael and Ester Vaida. 
This document was prepared by the SPT-3G collaboration using the resources of the Fermi National Accelerator Laboratory (Fermilab), a U.S. Department of Energy, Office of Science, Office of High Energy Physics HEP User Facility. Fermilab is managed by Fermi Forward Discovery Group, LLC, acting under Contract No. 89243024CSC000002.
The Melbourne authors acknowledge support from the Australian Research Council’s Discovery Project scheme (No. DP260100705). 
The Paris group has received funding from the European Research Council (ERC) under the European Union’s Horizon 2020 research and innovation program (grant agreement No 101001897), and funding from the Centre National d’Etudes Spatiales. 
The SLAC group is supported in part by the Department of Energy at SLAC National Accelerator Laboratory, under contract DE-AC02-76SF00515.
The CAPS authors are supported by the Center for AstroPhysical Surveys (CAPS) at the National Center for Supercomputing Applications (NCSA), University of Illinois Urbana-Champaign.
WLKW acknowledges support from an Early Career Research Award DE-SC0026376 of the Department of Energy.
This research was done using services provided by the OSG Consortium \citep{pordes07, sfiligoi09}, which is supported by the National Science Foundation awards 2030508 and 2323298.
The data analysis pipeline also uses the scientific python stack \citep{hunter07, jones01, vanDerWalt11}.
This work has made use of data from the European Space Agency (ESA) mission {\it Gaia} (\url{https://www.cosmos.esa.int/gaia}), processed by the {\it Gaia} Data Processing and Analysis Consortium (DPAC, \url{https://www.cosmos.esa.int/web/gaia/dpac/consortium}). Funding for the DPAC has been provided by national institutions, in particular the institutions participating in the {\it Gaia} Multilateral Agreement.

\bibliography{rs}

\appendix \label{sec:appA}
\setcounter{figure}{0}
\renewcommand{\thefigure}{A\arabic{figure}}
\setcounter{table}{0}
\renewcommand{\thetable}{A\arabic{table}}
\section{Data Processing}
\label{sec:app_maps}
\subsection{Mapmaking and Filtering}
\label{sec:filtering}
\begin{figure*}
\centering
    \includegraphics[width=0.95\linewidth]{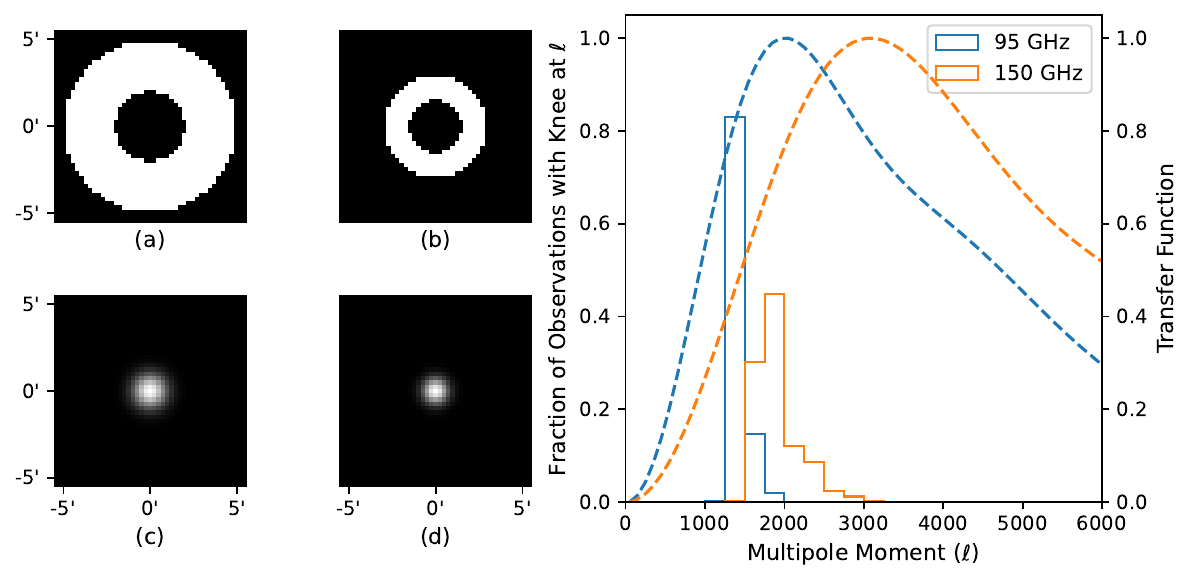}
    \caption{\label{fig:filters} Filter kernels and transfer function of the map-space filters used to process our difference maps. \textbf{Left, top row:} the annulus-shaped high-pass filter kernels used for the 95 GHz (a) and 150 GHz (b) maps. The map is convolved with the kernel to create a moving average, which is then subtracted from the input map. \textbf{Left, bottom row:} the Gaussian filter kernels used to maximize the signal-to-noise ratio for point sources. The kernels are the best-fit radially symmetric Gaussian approximation to the SPT-3G instrumental beam at 95 GHz (c) and 150 GHz (d). \textbf{Right:} histograms of the $1/\ell$-knee of SPT-3G observations, defined as the point where the noise power spectrum density reaches twice the value for the small-scale white noise. The dotted lines show the (radially averaged) transfer function of the combined HPF-and-beam filter, showing that the HPF (rising part) is reasonably well matched to the $1/\ell$-knee of the large-scale red noise for a wide variety of atmospheric conditions.}
\end{figure*}

We filter the time ordered data (TOD) produced by the detectors over the course of an observation and bin them into 2D pixelated maps by tracking the pointing for each detector. All maps used in this analysis use $0.25'$\,pixelization and a Lambert Azimuthal Equal-Area projection. We further filter the maps in order to suppress large-scale atmospheric fluctuations and maximize sensitivity to point sources. The TOD filtering is described in detail in \cite{quan26}, and the map-space filtering is described in \cite{tandoi24}. Here we provide a brief summary and note any pertinent changes. Filtering scales and functions are stated either in terms of spatial scale (arcmin) or angular multipole moment ($\ell$). 
The detector TOD have a bandpass filter applied with effective high- and low-pass cutoffs in the scan direction of $\ell=500$ and $\ell=20000$, respectively. From each map, we subtract a fixed sky template made from a 3-year average of Main field observations, and then apply two map-space filters. The first is an isotropic high-pass filter with an annulus-shaped kernel, where the outer radius is set to $5'\;(3')$ for \SI{95}{GHz}\;(\SI{150}{GHz}), respectively. 
The inner radius is set to three times the best-fit Gaussian standard deviation of the instrumental beams. 
For the  SPT-3G beams that have $\text{FWHM} = 1.57'\;(1.21')$ at 95 GHz (150 GHz), the inner radii are then $2.00'\;(1.54')$ at 95 GHz\;(150 GHz). The filter acts as a high-pass by subtracting the average of the pixels located between the inner and outer radii from a given pixel, removing noise at spatial scales larger than the outer radius. The dominant noise source at large spatial scales is thermal emission from the atmosphere. As weather conditions at the South Pole change from day to day, the amplitude of the atmospheric fluctuations can vary substantially, which we quantify by measuring the  ``1/$\ell$ knee'', or angular scale where the red atmospheric power spectrum equals the instrumental white noise floor. The shape of the high-pass filter is chosen to optimize point source detection in a wide variety of weather conditions. Figure \ref{fig:filters} shows the distribution of $1/\ell$-knees in our maps. These processing steps produce maps free of static signals and optimized for the detection of unresolved transient signals. 
Finally, we convolve the maps with a filter that is a Gaussian approximation of the instrument beam in each band and convert the map from differential temperature (\(\Delta T_{\text{CMB}}\)) to flux density in units of mJy. The combined transfer function of the two spatial filters is shown as the dashed lines in Figure \ref{fig:filters}. 

\subsection{Noise Estimation}
\begin{figure*}
    \centering
\includegraphics[width=0.95\linewidth]{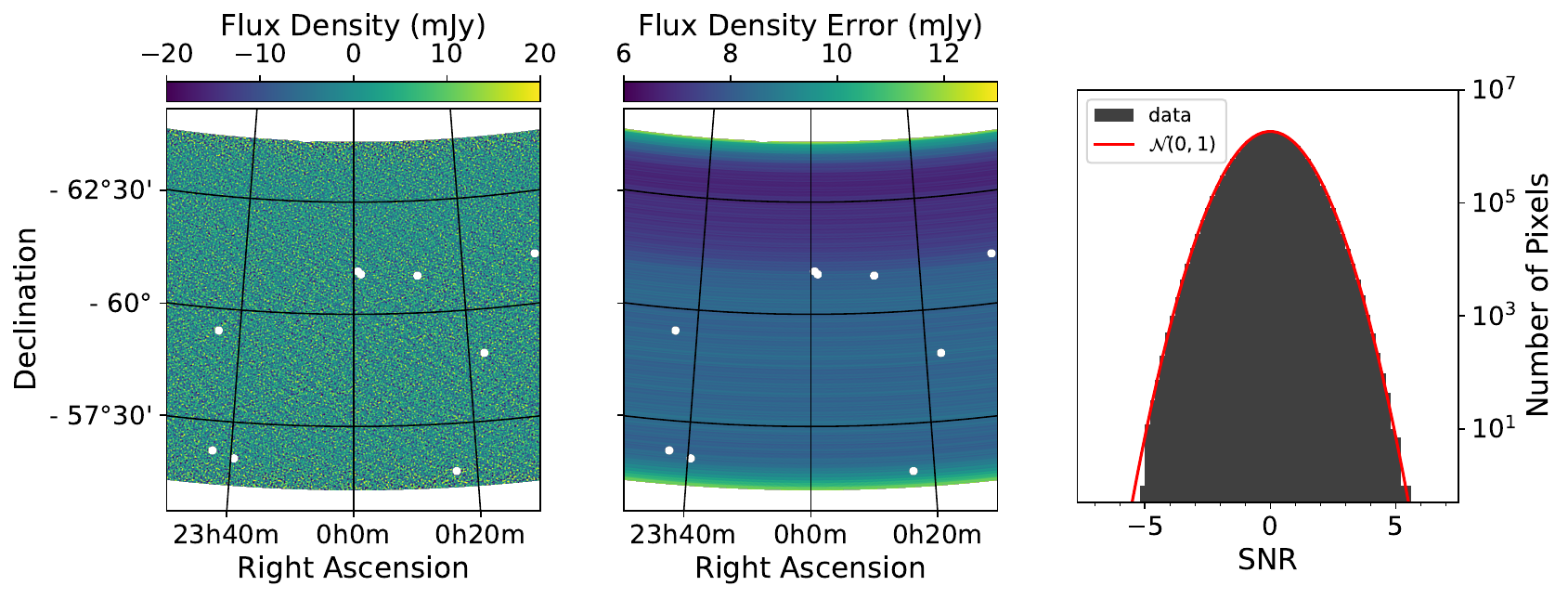}
    \caption{\label{fig:thumbnails} Filtered intensity map (left), error map (center), and signal-to-noise histogram (right) of an example SPT-3G map. After the filtering operations described in Section \ref{sec:filtering}, our maps are white-noise dominated. The corresponding error map shows the characteristic striped pattern created by holding the individual detector weights constant across telescope scans. The white disks are locations of very bright point sources, which are removed during the mapmaking step. After filtering the map and normalizing the inverse-variance weight map, the signal-to-noise map (formed by dividing the intensity map by the error map) is, to good approximation, described by a standard Gaussian $\mathcal{N}(0,1)$. }
\end{figure*}

During the map binning, we also produce a map of weights by calculating the average power spectral density (PSD) for each detector and summing the result into skymaps with the same pixelization as the intensity map. Because the detector PSDs are calculated across the entire width of a scan, the weight for any detector is constant across a single telescope scan. Our filtering pipeline produces white-noise dominated maps, for which the weight map $W$ is proportional to the inverse variance of the intensity map $T$. We fit an overall normalization by demanding that the pixels in a map of signal-to-noise ratio, $T\sqrt{W}$, are normally distributed with standard deviation equal to $1$. The normalization factor is fit to negative pixels only, to prevent bias from signals on the sky. We divide the weight map by this measured normalization factor in order to produce a true inverse-variance weight map. Figure \ref{fig:thumbnails} shows a large cutout of an example map, along with the corresponding error map and a full-map signal-to-noise histogram. A typical map will have noise variations on the order of 20\% across the map, with smaller variations of order 10\% on small scales due to variations in detector coverage. We show the median pixel noise for each map in Figure \ref{fig:noise_histo}. 

\subsection{Masking}
\label{sec:masking}
The SPT Main field contains thousands of point sources detected with high signal-to-noise. The vast majority can be roughly categorized as either rising-spectrum dusty star forming galaxies (DSFG) or falling-spectrum AGN. We mask a $4'$ radius area around each point source with a 3-year average 150 GHz flux of 5 mJy or brighter. Very bright sources ($>$ 50 mJy at 150 GHz) are interpolated over in the detector TOD to prevent large-scale filtering artifacts from contaminating the map. The total sky area removed by the point source mask is 34.4 square degrees. We also apply a weight-based boundary mask to each map, removing all points on the map edge where the pixel weights are below half the median map weight. Since the telescope scans are designed to overshoot the nominal RA bounds, the apodization mask does not remove any area outside the nominal SPT-3G Main survey field. In addition to the static point source mask, we also mask the locations of a small number of asteroids in our field using the asteroid tracking tools developed for \cite{chichura22}. 

\subsection{Map Cuts}
\label{sec:app_mapcuts}
\begin{figure*}[t]
    \centering
\includegraphics[width=0.5\textwidth]{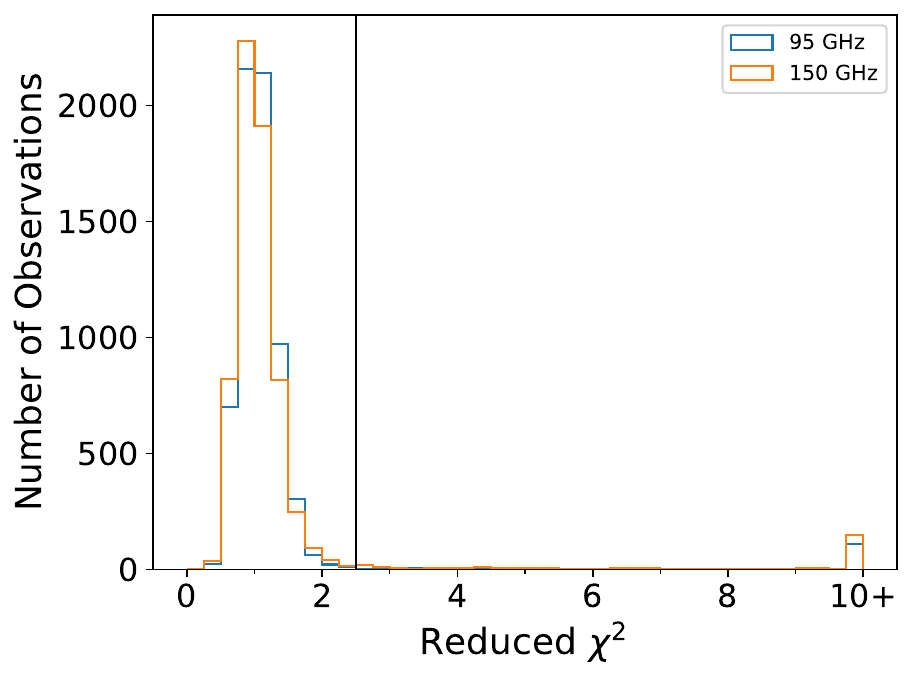}
    \caption{\label{fig:chi2} Distribution of the goodness-of-fit statistic testing the Gaussianity of single-observation map noise after map filtering and pixel noise estimating procedures. The statistic is the reduced $\chi^2$ of fitting a standard normal distribution to the negative pixel signal-to-noise histogram. All maps with $\chi^2 > 2.5$ are cut from the transient pipeline. }
\end{figure*}

Because the noise-normalization procedure and the transient-finding method rely on the map noise being Gaussian to good approximation, we measure the Gaussianity of each map and cut maps that show large deviations. Our signal of interest (transients) consists of arcminute-scale positive-amplitude fluctuations set in a white noise background covering many hundreds of square degrees, so we do not expect the presence of transient sources to significantly affect noise estimation. Still, to mitigate any possible signal-induced bias we opt for a binned goodness-of-fit metric where we measure a reduced $\chi^2$ limited to the negative side of the pixel SNR histogram of each map: 
\begin{equation}
    \chi^2_{\mathrm{r}} = \frac{1}{34} \sum_{i}^{35} \frac{\left(n_i - e_i\right)^2}{e_i}, 
\end{equation}
where $i$ runs over bins of width $0.1$ from SNR $=-3.5$ to SNR $=0$,~ $n_i$ is the measured number of pixels in each bin, and $e_i$ is the expected number of pixels for a standard Gaussian $\mathcal{N}(0,1)$ with mean 0 and variance 1. We sample every third pixel in both the $x$ and $y$ directions in order to suppress beam-scale correlations in the pixel values. The lower bound of $-3.5$ is chosen to ensure enough pixels populate each bin. Figure \ref{fig:chi2} shows histograms of the $\chi^2_r$ values for all maps in the dataset. Based on the histograms, we visually set a cutoff value of $\chi^2_r = 2.5$ at the edge of the bulk distribution and remove all maps above this value from our dataset. The resulting cut removes 2.4\% of maps at 95 GHz and 4.2\% of maps at 150 GHz. The cut maps show noise features consistent with excess atmospheric power leaking past the high pass filter and leaving residual large scale non-Gaussian noise in the data.

\subsection{Noise Renormalization}
\label{sec:app_renorm}
\begin{figure*}[t]
    \centering
\includegraphics[width=0.5\textwidth]{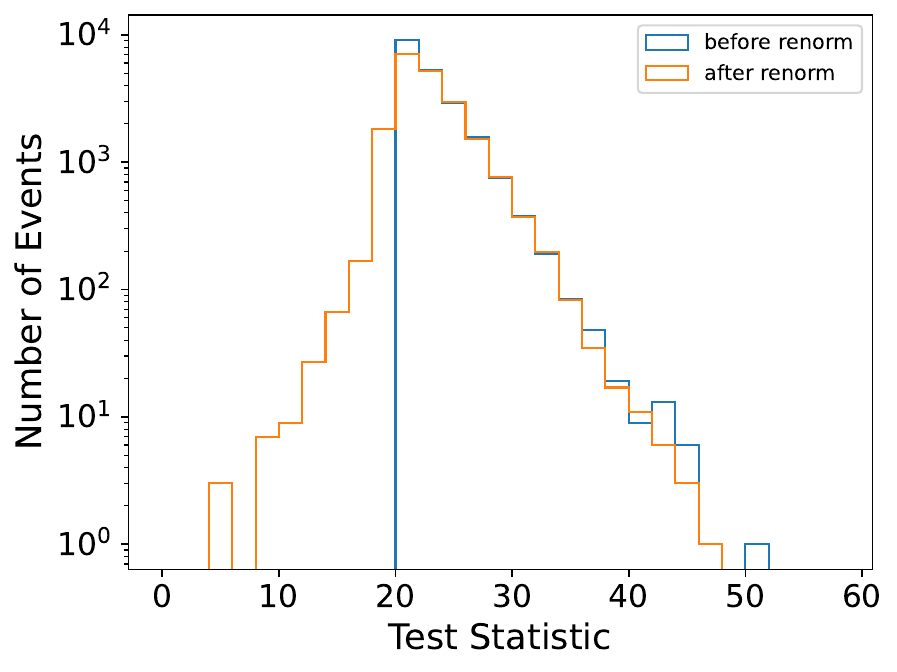}
    \caption{\label{appendix:renorm_fig} Background TS distribution before and after the noise renormalization procedure. The renormalization smears the initial TS distribution and removes a small number of high-significance false positives in the right tail. }
\end{figure*}

In some maps, leakage of excess atmospheric power occurs on short timescales and hence only affects a small part of the map. This results in a change of the effective normalization constant and a bias in the pixel noise estimates in the affected area. If the area is small, most of the map noise will be estimated correctly and the map will pass our cuts, 
potentially introducing false detections if the initial noise estimate in the affected area is too low, or causing real events to be missed if the estimate is too high. In practice, only the former occurs with any regularity. 
We currently mitigate this effect by applying a second noise estimation step to each transient event. For this renormalization step, we cut a \SI{1}{\degree} wide thumbnail around the source location, place a \SI{5}{\arcmin} radius mask at the center and recalculate the noise normalization constant using only pixels in the thumbnail. After correcting the noise estimate we then recalculate the TS of the event. This method is similar to the two-step normalization procedure described in the ACT transient search in \cite{biermann24}, with the major difference being that we renormalize the noise post-detection. We apply the procedure to every event with an initial $\mathrm{TS} \geq 20$. We compare the negative (background) event TS distributions before and after renormalization in Figure~\ref{appendix:renorm_fig}. The inherent variability between the global and local noise estimates is on the order of 5\%, and the main effect of the renormalization procedure is to smear the TS distribution by that amount. The initial cutoff leads to an incompleteness in the lowest few TS bins, as no $\mathrm{TS}<20$ events are present to be moved to higher values. Additionally, near the right edge of the distribution several ``high significance" events are removed by the procedure. Visual inspection indicates that these are indeed caused by streaks of elevated noise that were not properly accounted for by our initial weighting and normalization. Future transient searches could be improved by subtracting bright atmospheric fluctuations from the map directly, after estimating their amplitude and shape from the currently unused 220 GHz maps using the atmospheric mitigation techniques developed in \cite{coerver24}.

\setcounter{figure}{0}
\renewcommand{\thefigure}{B\arabic{figure}}
\setcounter{table}{0}
\renewcommand{\thetable}{B\arabic{table}}
\section{Flare Stars}
\label{sec:app_stars}
\begin{figure}
    \centering
    \includegraphics[width=\linewidth]{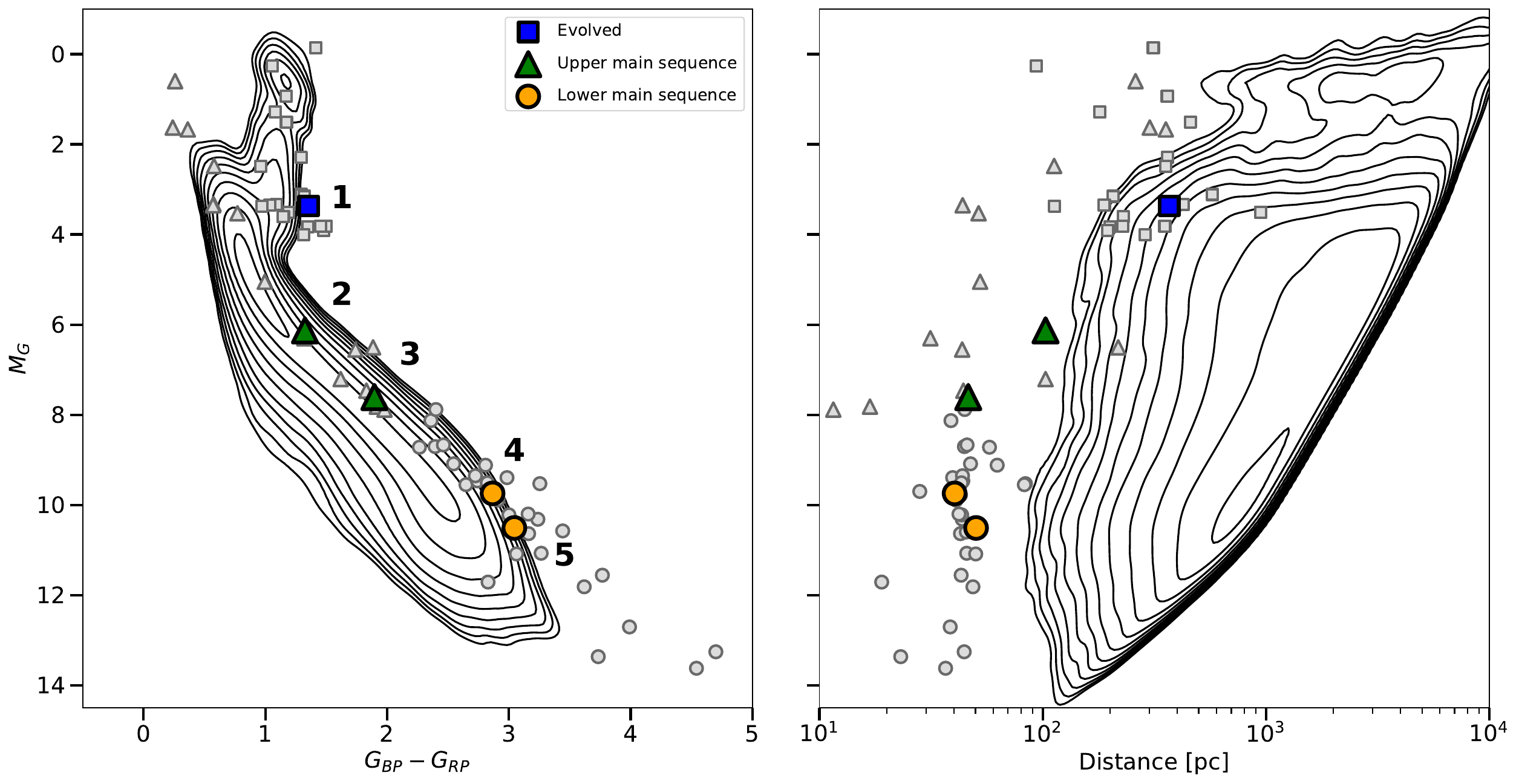} 
    \caption{\label{appendix:CMD_fig}
    Updated figure from \citet{tandoi24} showing flaring stars previously discovered by SPT-3G in gray and new detections from this work in color.
    \textbf{Left}: Color-magnitude diagram of SPT flaring stars, with contours showing density of Gaia stars in the SPT-3G winter field.
    \textbf{Right}: Magnitude vs distance for the same populations. 
    }
\end{figure}

This Appendix contains information on the five stellar flares found in this work that were not already reported in the 4-year SPT-3G stellar flare catalog \citep{tandoi24}. The flare time, counterpart, test statistic, and other data for each event are given in Table \ref{appendix:stars_table}. Four flares (numbers 1--4) are associated with stars not previously present in the catalog. The associated stars are of various types, but are overall consistent with the broader population in the published SPT-3G catalog. We show the stars on color-magnitude and magnitude-distance diagrams of the catalog population in Figure \ref{appendix:CMD_fig}.

\begin{table*}
\centering
\caption{\label{appendix:stars_table}Stellar flares found in this analysis but not previously reported in the 4-year SPT-3G stellar flare catalog. 
}
\fontsize{8}{9}\selectfont
\begin{tabular*}{0.99\textwidth}{@{\extracolsep{\fill}}lccccccc}
\toprule
\toprule
      \multicolumn{1}{c}{} &
      \multicolumn{1}{c}{\textit{Gaia} DR3} &
      \multicolumn{1}{c}{TS} &
      \multicolumn{1}{c}{p} &
      \multicolumn{1}{c}{MJD} &
      \multicolumn{1}{c}{\vtop{\hbox{\strut $\nu L_{\nu, 95}$}\hbox{\strut [erg s$^{-1}$]}}} &   
      \multicolumn{1}{c}{\vtop{\hbox{\strut $\nu L_{\nu, 150}$}\hbox{\strut [erg s$^{-1}$]}}} &   
      \multicolumn{1}{c}{$\alpha^{95}_{150}$} \\ 
\midrule
1 & 4752256797628777728 & 49.78 &$2.06\times10^{-2}$  & 58781.74 & 4.25$\times10^{29}$ & 1.51$\times10^{30}$ &  1.77$\pm$0.77\\
2 & 4721546166554166528 & 48.75 & $1.08\times10^{-3}$ & 58739.86 & 3.09$\times10^{28}$ & 7.97$\times10^{28}$ &  1.07$\pm$0.69\\
3 & 6409848126430574592 & 51.71 & $7.54\times10^{-3}$ & 59456.83 & 1.14$\times10^{28}$ & 1.69$\times10^{28}$ & -0.13$\pm$0.47\\
4 & 6503059740071544064 & 47.17 & $1.37\times10^{-2}$ & 59360.40 & 6.20$\times10^{27}$ & 6.21$\times10^{27}$ & -0.99$\pm$0.94\\
5 & 6456528236185330816 & 49.61 & $4.44\times10^{-3}$ & 59083.43 & 6.75$\times10^{27}$ & 1.05$\times10^{28}$ & -0.03$\pm$0.98\\
\bottomrule
\end{tabular*}
\tablenotetext{}{The spectral index $\alpha^{95}_{150}$ ($S_\nu\propto \nu^\alpha$) and association p-value are as defined in \cite{tandoi24}.}
\end{table*}

\end{document}